\documentclass[final]{siamltex}

\usepackage{amsmath}
\usepackage{amssymb}
\usepackage{tikz}
\usepackage{graphicx}
\usepackage{cite}
\usepackage{subfig}
\usepackage{color}
\usepackage{rotating}
\usepackage{mathtools}
\usepackage{xcolor}
\usepackage{showkeys}
\usepackage{pifont}

\usetikzlibrary{decorations.pathmorphing}
\usetikzlibrary{shapes}
\usetikzlibrary{plotmarks}

\renewcommand{\inf}{\infty}
\newcommand{\eps}{\epsilon}

\newcommand{\e}{\mathrm{e}}
\renewcommand{\d}{\,\mathrm{d}}
\newcommand{\diff}[2]{\frac{\mathrm{d} #1}{\mathrm{d} #2}}

\def\XXint#1#2#3{{\setbox0=\hbox{$#1{#2#3}{\int}$}
\vcenter{\hbox{$#2#3$}}\kern-.5\wd0}}

\usepackage{setspace}
\title{Nanoptera in a Period-2 Toda Chain
}

\author{
Christopher J. Lustri\footnotemark[1]
\and Mason A. Porter\footnotemark[2]
}

\begin{document}

\maketitle

	\renewcommand{\thefootnote}{\fnsymbol{footnote}}
	\footnotetext[1]{Department of Mathematics, Macquarie University, Sydney, Australia (christopher.lustri@mq.edu.au).}
	\footnotetext[2]{Oxford Centre for Industrial and Applied Mathematics, Mathematical Institute, Oxford OX2 6GG, UK; CABDyN Complexity Centre, University of Oxford, Oxford OX1 1HP, UK; and Department of Mathematics, University of California, Los Angeles, Los Angeles, California 90095, USA (mason@math.ucla.edu).}
		
	\renewcommand{\thefootnote}{\arabic{footnote}}


\begin{abstract}
We study asymptotic solutions to a singularly-perturbed, period-2 Toda lattice and use exponential asymptotics to examine `nanoptera',  which are nonlocal solitary waves with constant-amplitude, exponentially small wave trains. With this approach, we isolate the exponentially small, constant-amplitude waves, and we elucidate the dynamics of these waves in terms of the Stokes phenomenon. We find a simple asymptotic expression for the waves, and we study configurations in which these waves vanish, producing localized solitary-wave solutions. In the limit of small mass ratio, we derive a simple anti-resonance condition for the manifestation these wave-free solutions.
\end{abstract}







\section{Introduction}\label{intro}

In this paper, we examine the dynamics of a spatially-localized wave propagating along a chain of particles, where the dynamics of each particle is governed by an interaction potential between it and its nearest neighbors. Systems of this form have been studied with a variety of different interaction potentials---including in celebrated models such as the Toda lattice \cite{Toda1} and the Fermi--Pasta--Ulam--Tsingou (FPUT) problem \cite{FPU,fpupop} and in experimental systems such as granular crystals (via Hertzian interactions)\cite{Nesterenko1,Sen1,PT2015,chong2017} and their generalizations (see, e.g., \cite{yang2014,bonanomi2015,gantz2013}), chains of magnets \cite{magnet2014}, and more.

Classical examples of lattice systems such as the Toda lattice have been studied extensively both because of their intrinsic theoretical interest and because they play important roles in several applications, including nonlinear optical propagation \cite{Arnold1} and electrical propagation through transmission lines \cite{Hirota1,Kuusela1}. The Toda-lattice equations also provide a paradigmatic example of a discrete integrable system. Consequently, properties associated with the integrability of the Toda-lattice equations have been the subject of numerous studies. In particular, the integrability of the Toda lattice allows it to support soliton solutions \cite{Toda1}. 

Numerous studies have considered `uniform' (i.e., monoatomic) one-dimensional (1D) lattices, which are chains in which each particle in the lattice is identical. However, particles need not be identical, and examinations of such `heterogenous' lattices \cite{Flach:ArxivRep2014-2}, with either periodic \cite{Jayaprakash1,Jayaprakash2,Porter1,Vainchtein1} or random \cite{Flach:ArxivRep2014-1,jandro2015} distributions of different particles, reveal a wealth of fascinating dynamics that do not arise in uniform lattices. Such dynamics include new families of solitary waves that have been observed in diatomic granular chains and which can exist only for discrete values of the ratio between the masses of the two particles in a diatomic unit \cite{Jayaprakash1}.  

\begin{figure}
\centering
\begin{tikzpicture}
[xscale=1,>=stealth,yscale=1]

\draw [line width=0.3mm] (-1.5,0) node[left] {$\ldots$} -- (3.25,0) node[right] {$\ldots$};
\fill [white] (-1,0) circle (0.25);
\fill [white] (-0.25,0) circle (0.25);
\fill [white] (0.5,0) circle (0.25);
\fill [white] (1.25,0) circle (0.25);
\fill [white] (2,0) circle (0.25);
\fill [white] (2.75,0) circle (0.25);
\draw (-1,0) circle (0.25);
\draw (-0.25,0) circle (0.25);
\draw (0.5,0) circle (0.25);
\draw (1.25,0) circle (0.25);
\draw (2,0) circle (0.25);
\draw (2.75,0) circle (0.25);
\node at (0.5,0.35) [above] {\scriptsize{$n-1$}};
\node at (1.25,0.375) [above] {\scriptsize{$n$}};
\node at (2,0.35) [above] {\scriptsize{$n+1$}};
\draw (1.25,0) -- (1.25,-0.9) node[below] {\scriptsize{$q(n,t)$}};
\filldraw[black] (1.25,0) circle (0.05);
\draw [<->] (1,-0.6) -- (1.5,-0.6);
\node at (0.875,1.5) {\scriptsize{(a) Particles with constant mass}};

\draw [line width=0.3mm] (5.5,0) node[left] {$\ldots$} -- (10.25,0) node[right] {$\ldots$};
\fill [white] (6,0) circle (0.25);
\fill [white] (6.75,0) circle (0.25);
\fill [gray,opacity=0.25] (6.75,0) circle (0.25);
\fill [white] (7.5,0) circle (0.25);
\fill [white] (8.25,0) circle (0.25);
\fill [gray,opacity=0.25] (8.25,0) circle (0.25);
\fill [white] (9,0) circle (0.25);
\fill [white] (9.75,0) circle (0.25);
\fill [gray,opacity=0.25] (9.75,0) circle (0.25);
\draw (6,0) circle (0.25);
\draw (6.75,0) circle (0.25);
\draw (7.5,0) circle (0.25);
\draw (8.25,0) circle (0.25);
\draw (9,0) circle (0.25);
\draw (9.75,0) circle (0.25);
\node at (7.875,1.5) {\scriptsize{(b) Particles with alternating masses}};

\draw (6.25,-1) -- (6.5,-1) -- (6.5,-0.875) node [right] {\scriptsize{: Particle of mass $m_1$}} -- (6.5,-0.75) -- (6.25,-0.75) -- cycle;
\fill [gray,opacity=0.25] (6.25,-1.25) -- (6.5,-1.25) -- (6.5,-1.5) -- (6.25,-1.5) -- cycle;
\draw (6.25,-1.25) -- (6.5,-1.25) -- (6.5,-1.375) node [right] {\scriptsize{: Particle of mass $m_2$}} -- (6.5,-1.5) -- (6.25,-1.5) -- cycle;

\end{tikzpicture}

\caption{(a) A uniform particle chain. The quantity $q(n,t)$ represents the horizontal displacement of the particle in position $n$ from its equilibrium position at time $t$. (b) A period-2 particle chain that consists of alternating particles with different masses.}\label{F:IntroPic}
\end{figure}
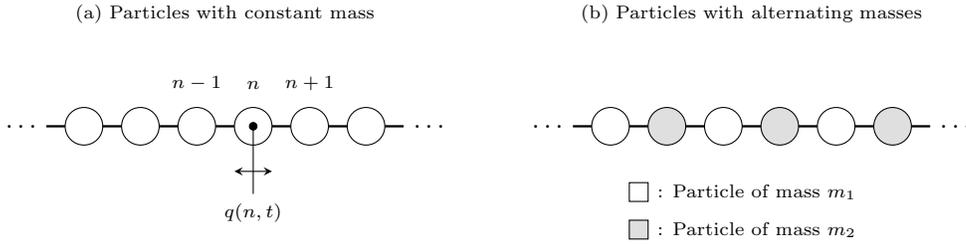

In the present study, we apply asymptotic techniques to consider the behavior of propagating waves in diatomic (i.e., `period-2') lattices, in which two different types of structures alternate with each other. Specifically, we consider bidisperse lattice systems, in which there are two different types of particles, that consist of alternations between a single heavy particle and a single light particle. See Figure \ref{F:IntroPic} for a schematic. The primary purpose of our paper is to investigate the behavior of solitary waves in the period-2 Toda lattice in the limit in which the ratio between the heavy and light particles becomes small (i.e., $m_2/m_1 \rightarrow 0$). Specifically, motivated by previous mathematical and experimental studies \cite{Okada1, Kofane1,Vainchtein1}, we examine solutions to these period-2 Toda lattice systems that demonstrate `nonlocal' solitary waves (so-called `nanoptera'), which approximately satisfy the classical definition of a spatially localized solitary wave, with the addition of small-amplitude, non-decaying oscillations on one or both sides of the wave front. 

The diatomic Toda lattice was studied by Vainchtein et al. \cite{Vainchtein1}, who used scale separation and the integrability of the system to obtain an asymptotic expression for the correction to a monoatomic soliton solution in terms of hypergeometric functions (see Eq.~\eqref{4:exact}). Their study conjectured that there exist an infinite number of parameter combinations with nanopteron-free solutions. These parameter combinations correspond to `anti-resonance' conditions \cite{Jayaprakash1, Jayaprakash2}. In the present investigation, we obtain the asymptotic behavior of the wave train using exponential-asymptotic methods, rather than by using scale separation. This approach permits us to directly study the dynamics of nanoptera in the solution without calculating any terms beyond the leading-order soliton solution. We use these results to obtain a simple expression for the anti-resonance condition in the asymptotic limit that is consistent with the conjecture made in \cite{Vainchtein1}. 


\subsection{Diatomic Chains}\label{diatomic}

One of the earliest investigations of the 2-periodic Toda lattice was a numerical study of the near-integrable properties of the system by Casati and Ford \cite{Casati1}. Their study showed that the diatomic system behaves as a near-integrable system for values of the mass ratio $m_2/m_1 \in (0,1)$ (where we note that the system is integrable for $m_2/m_1 = 0$ and $m_2/m_1 = 1$) if the total system energy is below some critical threshold, which depends on the mass ratio. If the total energy of the system exceeds this threshold, the system transitions from near-integrable to chaotic dynamics. In our study, we examine the near-integrable regime of the system, as we only consider small mass ratios ($0 < m_2/m_1 \ll 1$). The threshold energy becomes infinite as the system nears the integrable dynamics associated with $m_2/m_1 = 0$.

The knowledge that the diatomic Toda lattice exhibits near-integrable dynamics has inspired many authors to search for solitary-wave solutions to these periodic systems. Early results established that these lattices do not produce the localized pulse-like solutions that are typically associated with the Toda Lattice \cite{Toda2}. Nevertheless, numerical studies such as \cite{Hornquist1, Okada1, Tabata1}, as well as experimental studies using electrical transmission lines \cite{Kofane1}, indicate that any solitary wave that propagates through a diatomic Toda lattice leaves a trail of non-decaying oscillations in its wake. In particular, the work of Okada et al. \cite{Okada1} suggested that the amplitude of these oscillations decays exponentially in the small mass-ratio limit, so such a system is a natural one to analyze using exponential asymptotic methods. Recently, the study by Vainchtein et al. \cite{Vainchtein1} (which we discussed briefly in Section \ref{intro}), used a scale-separation method to obtain an asymptotic solution to the diatomic Toda system in terms of hypergeometric functions.

The earliest analytical work on the behavior of solitary waves in a diatomic Toda lattice system was that of Dash and Patanaik \cite{Dash1}, who extended the work of Toda on the uniform lattice to identify solitary-wave dynamics. Their solutions were complex-valued and hence not physical \cite{Mokross2}, but their results suggested that analytical methods can be adapted to the diatomic lattices. 

Several studies have examined the dynamics of solitary waves in a variety of different diatomic lattices, including systems with cubic and quartic nearest-neighbor potentials \cite{Dash2,Pnevmatikos1,Yajima1}, Lenard--Jones potentials \cite{Cassatella1}, Klein--Gordon lattice chains \cite{Kivshar1}, and Fermi--Pasta--Ulam systems \cite{Qin1}. An important example for engineering applications is diatomic granular crystals (with alternating heavy and light particles) \cite{chong2017}, which have been studied both experimentally \cite{Porter1} and numerically and asymptotically \cite{Jayaprakash1, Jayaprakash2, Porter1}. Granular crystals have Hertzian interactions between neighboring particles, and they exhibit fascinating resonant behavior for discrete sets of mass ratios \cite{Jayaprakash1}. These resonances produce discrete families of solitary waves with monotonically decaying tails, even though Hertzian chains generically produce nonlocal solitary-wave solutions. Similar phenomena have also been studied both numerically and analytically for a variety of `mass-in-mass' periodic particle chains \cite{Kevrekidis1, Liu2, Xu1}, and we will discuss such phenomena briefly in the context of the period-2 Toda lattice in Section \ref{S:2.3}. There have also been numerous studies of the thermal properties of period-2 Toda lattices (see, e.g., \cite{Diederich1, Hatano1, Mokross1}), and such properties have also been studied in other diatomic particle systems, including the Fermi--Pasta--Ulam-$\beta$ lattice \cite{Zhang3} and systems with periodic nearest-neighbor potentials \cite{Savin1}. 

Previous studies of particle chains have often utilized long-wave continuous approximations to find approximate solitary-wave solutions \cite{Dash2,Pnevmatikos1,Yajima1}. Other studies of discrete chains have taken the alternative approach of expressing exact solutions in terms of integral equations, and \cite{Ahnert1} used this idea to show that a general class of monoatomic Hamiltonian lattices with power-law interactions possess localized solitary-wave solutions that decay superexponentially in space. Several studies of diatomic particle chains, such as \cite{Cassatella1, Jayaprakash1, Jayaprakash2}, have applied asymptotic techniques directly to the governing differential--difference equations (rather than to integral equations).

In the present article, we examine nanoptera in a diatomic Toda chain. The singular nature of perturbing from a monoatomic chain (which has a mass ratio of $1$) to a diatomic chain suggests that it is helpful to use exponential asymptotic techniques to asymptotically investigate the dynamics of waves that propagate through diatomic particle chains. This is also true for systems that are perturbations of a chain with mass ratio $0$, such as the ones that we consider in our study (see Eqs,~\eqref{0:2P1},\eqref{0:2P2}). Specifically, one should expect to observe a form of nonlocal solitary wave known as a `nanopteron' (see Section \ref{nano} for a detailed description), in which non-decaying trains of small waves appear across curves known as `Stokes curves'. The existence of these waves in the period-2 Toda lattice was suggested by numerical calculations in \cite{Okada1} and by experimental observations in \cite{Kofane1}, and it was subsequently studied using multiple-scale asymptotics in \cite{Vainchtein1}. In the present study, we derive asymptotic descriptions of these waves using exponential asymptotic methods.


\subsection{Nanoptera}\label{nano}

J.~P. Boyd introduced the concept of a `nanopteron' in his study of the $\phi^4$ breather \cite{Boyd4}. Boyd used the term to describe `weakly nonlocal solitary waves', which approximately satisfy the classical definition of a solitary wave, except that a nanopteron wave asymptotes to a small-amplitude oscillation on either one (`one-sided') or both (`two-sided') sides of the central solitary wave. That is, the difference between a traditional solitary wave and a nanopteron is that the latter are localized spatially only up to these small oscillations of non-decaying amplitude. We emphasize that `nonlocal solitary wave' does not refer to nonlocal dependence, such as an integral equation, but rather that the wave is not spatially localized, as in the classical definition of a solitary wave.

\begin{figure}
\centering
\subfloat[Solitary Wave]{
\begin{tikzpicture}
[xscale=0.4,>=stealth,yscale=2]
\draw[black,thick] plot[smooth] file {Sol1b.txt}; 
\draw (-5,1.2) -- (5,1.2) -- (5,-0.2) -- (-5,-0.2) -- cycle;
\draw [->] (1,0.5) -- (2,0.5) node[above right] {\scriptsize{$x = ct$}};
\end{tikzpicture}
}
\subfloat[One-sided nanopteron]{
\begin{tikzpicture}
[xscale=0.4,>=stealth,yscale=2]
\draw[black,thick] plot[smooth] file {Sol2b.txt}; 
\draw (-5,1.2) -- (5,1.2) -- (5,-0.2) -- (-5,-0.2) -- cycle;
\draw [->] (1,0.5) -- (2,0.5) node[above right] {\scriptsize{$x = ct$}};
\end{tikzpicture}
}
\subfloat[Two-sided nanopteron]{
\begin{tikzpicture}
[xscale=0.4,>=stealth,yscale=2]
\draw[black,thick] plot[smooth] file {Sol3b.txt}; 
\draw (-5,1.2) -- (5,1.2) -- (5,-0.2) -- (-5,-0.2) -- cycle;
\draw [->] (1,0.5) -- (2,0.5) node[above right] {\scriptsize{$x =ct$}};
\end{tikzpicture}
}
\caption{Comparison of the profiles associated with (a) a standard solitary wave, (b) a one-sided nanopteron, and (c) a two-sided nanopteron that each propagate at speed $c$. The solitary wave is localized spatially, whereas the nanoptera have non-decaying oscillatory tails on (b) one side or (c) both sides of the wave front.}\label{F:nanoptera}
\end{figure}
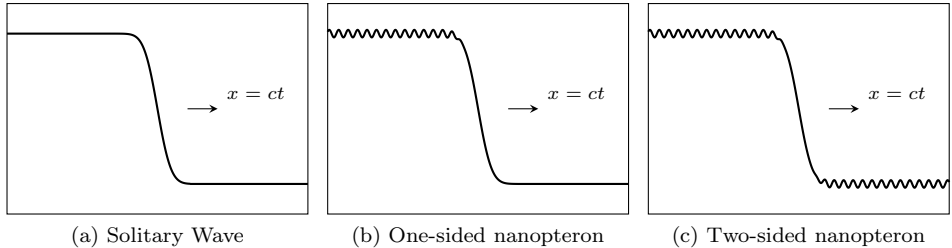

Typically, these oscillations (which, in some papers, are themselves called nanoptera) have exponentially small amplitude in the limit of some small parameter in a system; this causes the oscillations to be invisible to standard asymptotic power-series asymptotic methods. To analyze these oscillations asymptotically, it is useful to use exponential asymptotic techniques. In Figure \ref{F:nanoptera}, we illustrate different types of nanoptera and compare them to standard solitary waves.

Exponential asymptotic techniques, such as those developed in \cite{Berry1, Berry4, Chapman1, Daalhuis1}, can be used to capture the exponentially small oscillations associated with nanoptera. These techniques have been used previously to find nanopteron solutions to a variety of singularly-perturbed continuous systems---including the fifth-order Korteweg--de Vries equation for gravity-capillary waves \cite{Grimshaw2,Grimshaw1,Pomeau1,Trinh4}, a nonlinear Schr\"{o}dinger equation \cite{Wai1}, the $\phi^4$ Klein--Gordon equation \cite{Segur2}, and internal gravity waves \cite{Akylas2}. 

Alfimov et al. \cite{Alfimov1} used exponential asymptotic techniques from \cite{Boyd5} to study several systems with
exponentially small oscillatory waves. These examples include the discrete Klein--Gordon equation, which they studied in the continuum limit. Alfimov et al. \cite{Alfimov1} identified an orthogonality condition that causes the oscillatory waves to disappear precisely, leading to localized traveling-wave solutions. In the present study, we explicitly identify a similar orthogonality condition for the periodic Toda lattice in the small mass-ratio limit and explain how this condition arises from destructive interference between contributions that are switched on across different Stokes curves in the solution. 

More recently, exponential asymptotics have been used to study differential--difference systems. These techniques were used by Melvin et al. \cite{Melvin1} to investigate solutions to the 1D Salerno model and by Oxtoby and Barashenkov \cite{Oxtoby1} to study solutions to a discrete nonlinear Schr\"{o}dinger equation. In both of these studies, the authors found that traveling-wave solutions generally took the form of nanopterons, but they also identified localized traveling-wave solutions that arise as a consequence of the Stokes constant vanishing for particular parameter values. In the present study of the period-2 Toda lattice, we also find nanopteron-free solutions.\footnote{When we refer to solutions as `nanopteron-free', we are using the term to refer only to the small waves, in contrast to our usual convention of referring to the entire solution as a nanopteron.} However, in our case, these solutions are caused by the satisfaction of anti-resonance conditions, rather than by vanishing Stokes coefficients. Additionally, \cite{Melvin1} and \cite{Oxtoby1} used Borel transform methods for exponential asymptotics, whereas we apply the late-order-terms technique of Chapman et al. \cite{Chapman1}. Methods for exponential asymptotic methods, such as Borel transform methods \cite{Berry1, Berry4} and late-order-term methods \cite{Chapman1, Daalhuis1}, that were developed originally for studying ordinary differential equations can also be applied directly to many differential--difference equations, including the diatomic Toda lattice that is the subject of the present study. We discuss the reason for this at the beginning of Section \ref{S:2.2}.

In other diatomic lattices, such as the diatomic FPUT lattice \cite{Faver1, Hoffman1} and other periodic FPUT lattices \cite{Gaison1}, the existence of nanoptera has been established rigorously. Using a Beale ansatz, the authors of these papers decomposed a solution into a solitary-wave profile, a periodic component (the oscillatory wave train that is characteristic of nanoptera), and a small localized remainder. They obtained equations for the periodic component and the remainder; and, by controlling error estimates, the authors were able to prove the existence of nanopteron solutions. In \cite{Iooss1}, the authors applied similar techniques to a system of coupled nonlinear oscillators with both FPUT interactions and a potential term. They proved the existence of nanoptera with wave trains that are exponentially small with respect to a particular bifurcation parameter. 

In the present work, we do not perform a rigorous existence proof, and we instead take a similar approach to \cite{Vainchtein1} and compare the results of our formal analysis to computational results. We note that it would be an interesting problem to establish rigorous existence results for the diatomic Toda lattice, either using estimates on equations derived from a Beale ansatz, as in \cite{Faver1, Gaison1,Hoffman1}, or by using Borel-transform analysis as in \cite{Melvin1,Oxtoby1}. In this study, we use exponential asymptotic techniques developed by \cite{Chapman1, Daalhuis1} to provide a mathematical description of nanopteron solutions in the period-2 Toda lattice. We perform all asymptotic analysis on the spatially-discrete system (i.e., without using any continuum or quasi-continuum approximations). 

The rest of our paper is organized as follows. We discuss the method of exponential asymptotics in Section \ref{Method}. We use exponential asymptotics to analyze nanopteron solutions in a diatomic Toda lattice in Section \ref{S:2}. We first present general equations for chains with nearest-neighbor interactions in Section \ref{nearest}, and we then perform a full exponential asymptotic analysis in Sections \ref{S:2.1}--\ref{S:2.3} (with some of the technical details presented in Appendix \ref{S:2.2.5.2}). In Section \ref{S:Results}, we compare our asymptotic results to results from Vainchtein et al. \cite{Vainchtein1}. We conclude in Section \ref{S:Final}.


\section{Exponential Asymptotics and Stokes Curves}\label{Method}

We examine the asymptotic behavior of exponentially small, non-decaying waves that appear in the wake of a solitary-wave front in diatomic Toda lattices. However, determining the behavior of terms that are exponentially small compared to the leading-order solution in the $\eps \rightarrow 0$ asymptotic limit is impossible using classical asymptotic series expansions, because the exponentially small contribution is necessarily smaller than any power of the small parameter $\eps$. Therefore, we apply specialized techniques, known as \emph{exponential asymptotics}, to determine behavior on this scale \cite{Boyd1,Boyd3}.

As we uncover this exponentially small behavior, we will see that the analytic continuation of the solution includes curves known as `Stokes curves' \cite{Stokes1}. These curves are related to the behavior of exponentially small components of the solutions. As a Stokes curve is crossed, the exponentially small contribution experiences a smooth, rapid change in value in the neighborhood of the curve. In many problems, including the present investigation, the exponentially small contribution to the solution appears only on one side of a Stokes curve.

The central idea of exponential asymptotic methods is that one can truncate a divergent asymptotic series to provide a useful approximation to an exact solution. Additionally, one can choose the truncation point to minimize the error between the approximation and the exact solution; this is known as \textit{optimal truncation}. Importantly, when a divergent series is truncated optimally, the associated approximation error is generally exponentially small in the asymptotic limit \cite{Boyd1}. One can then rescale the problem to directly determine this approximation error, allowing the exponentially small component of the solution to be determined in the absence of the asymptotic series itself. This idea was introduced by Berry \cite{Berry1,Berry4}, and it was employed in \cite{Berry3,Berry5} to determine the position of Stokes curves in special functions such as the Airy function and to examine the smooth nature of the associated switching behavior. 

In the present paper, we apply an exponential asymptotic method developed by Olde Daalhuis et al. \cite{Daalhuis1} for linear differential equations and extended by Chapman et al. \cite{Chapman1} to nonlinear ordinary differential equations. We provide a brief outline of the process; see the above papers for a more detailed explanation of the methodology.

The first step in exponential asymptotic analysis is to express the solution as an asymptotic power series. In many singular perturbation problems, including those that we consider in the present study, the asymptotic series solution takes the form of a divergent series. For a more detailed discussion of asymptotic series divergence, see \cite{Boyd2,Dingle1}. One can minimize the approximation error by truncating the series optimally. However, this requires a general form for the asymptotic series coefficients, and it often is algebraically intractable to obtain such a general form. In practice, however, one does not require the exact form of the series coefficients. Instead, one needs only the so-called \textit{late-order terms}, which are asymptotic expressions for the $r$th series coefficient in the $r \rightarrow \inf$ limit.

For singular perturbation problems, Darboux \cite{darboux1878} noted that one can obtain successive terms in an asymptotic series expansion by repeated differentiation of an earlier term in the series. Specifically, assume that we have some function $f(z)$ on $z \in \mathbb{C}$ that can be expanded within a circle as a Taylor--Maclaurin series $\sum_0^{\inf} g_r z^r$. If we let $z_s$ represent the location of the nearest non-essential singularity on or outside the circle of convergence, we can write
\begin{equation}
	f(z) = (z_s - z)^{-p_s}f_s(z)\,,
\end{equation}
where $p_s$ is either fractional (for branch points) or a positive integer (for poles) and $f_s(z)$ can be expanded as a Taylor series about $z_s$. Darboux \cite{darboux1878} showed that the behavior of $g_r$ in the limit that $r \rightarrow \inf$ follows a predictable form, satisfying
\begin{equation}\label{result-asym}
	g_r \sim \frac{f_s(z_s)\Gamma(r+p_s)}{\Gamma(r+1)(p_s-1)z_s^{r+p_s}}\qquad \mathrm{as}\qquad r \rightarrow \inf\,,
\end{equation}
where $\Gamma$ is the gamma function \cite{DLMF, Abramowitz1,OLBC10}. Dingle \cite{Dingle1} gave a detailed derivation of the expression \eqref{result-asym}. Dingle also considered the case in which there are multiple equidistant singularities and demonstrated by example that their contributions can be summed to correctly approximate the late-order terms of the power-series expansion. This is also true of the behavior that we consider in the present study. 

From \eqref{result-asym}, we see that the late-order terms of the resulting asymptotic series diverge as the ratio between a factorial and the increasing power of a function $\chi$. Noting the factorial-over-power form in \eqref{result-asym}, Chapman et al. \cite{Chapman1} proposed writing an ansatz for the late-order terms that is capable of describing this form of late-order term behavior. One writes
\begin{equation}\label{ch1:ansatz}
	g_r \sim \frac{G\,\Gamma(r+\gamma)}{\chi^{r+\gamma}} \qquad \mathrm{as} \quad r \rightarrow \inf\,,
\end{equation}
where $G$, $\gamma$, and $\chi$ are functions that do not depend on $r$ but are free to vary with independent variables (and hence with $z$). The `singulant' $\chi$ equals $0$ at values of $z$, denoted by $z = z_s$, at which the leading-order behavior $g_0$ is singular. This ensures that the late-order ansatz for $g_r$ also has a singularity at $z = z_s$ and that the singularity increases in strength as $r$ increases. One can then use the ansatz \eqref{ch1:ansatz} to optimally truncate an asymptotic expansion. The method developed in Olde Daalhuis et al. \cite{Daalhuis1} involves substituting the resulting truncated series expression into the original problem to obtain an equation for the exponentially small remainder term. 

The exponentially small contribution to the asymptotic solution that one obtains using the above method generally takes the form $\mathcal{S}A\e^{-\chi/\eps}$, where the `Stokes multiplier' $\mathcal{S}$ varies rapidly from $0$ to a nonzero value as one crosses a Stokes curve. This behavior is known as \textit{Stokes switching}, and it occurs along curves at which the switching exponential is maximally subdominant and hence where the singulant $\chi$ is real and positive \cite{Berry5}. The variation is smooth, and it occurs in a neighborhood of width $\mathcal{O}(\sqrt{\eps})$ that surrounds the curve.

Examining when a Stokes multiplier becomes nonzero provides a simple criterion to determine where the exponentially small contribution to a solution `switches on' and cannot be ignored. One finds the positions of Stokes curves by determining the curves along which the singulant is real and positive. Consequently, exponential asymptotic analysis makes it possible to (1) determine the form of exponentially small contributions to a solution and (2) determine the regions of a solution domain in which the contributions are present. In our study, these contributions take the form of an exponentially small wave train that follows the leading-order soliton. In other words, they give us nanoptera.

We also briefly note the existence of another class of curve, known as `anti-Stokes curves'. These curves occur when exponential behavior changes between being exponentially small and exponentially large in some parameter, and they arise when the imaginary part of $\chi$ is $0$. We will see anti-Stokes curves in the present study, but they do not play a significant role in our conclusions.

This primary advantage of this approach is that it does not require the computation of terms beyond the leading-order expression to obtain the exponentially small correction terms. The leading-order behavior is required to determine the location of the singularities that are required to obtain $\chi$, but it is not even necessary to exactly calculation of the first correction term $g_1$. This makes the technique particularly useful for the many nonlinear problems for which obtaining even these low-order correction terms is intractable. See the review article \cite{Boyd1} or monograph \cite{Boyd3} for more details on exponential asymptotics and their application to nonlocal solitary waves, \cite{Berry1,Berry4,Boyd2} for examples of previous studies of exponential asymptotics, and \cite{Chapman1,Daalhuis1} for more details on the particular methodology that we apply in the present paper.


\section{Period-2 Toda Lattice}\label{S:2}

\subsection{Nearest-Neighbor Interactions in a Particle Chain} \label{nearest}

We consider chains of particles in which the dynamics of each particle is governed by nearest-neighbor interactions, with an interaction potential $\phi$. We illustrate an example of such a system in Figure \ref{F:IntroPic}(a).  The equations that govern the particle displacement of these chains is of the form
\begin{equation}\label{0:G}
	m \ddot{q}(p,t) = \phi'\left[q(p,t) - q(p-1,t)\right] - \phi'\left[q(p+1,t) - q(p,t)\right]\,,
\end{equation}
where $q(p,t)$ is the displacement from equilibrium at time $t$ of the particle located at position $p$, the mass of each particle is $m$, and \,$\dot{}$\, indicates a derivative with respect to time.  We also require boundary conditions at $t = 0$ and as $n \rightarrow \inf$. We will make these conditions explicit for the case of diatomic chains.

There have been many studies of systems of the form \eqref{0:G} that ask whether the equations admit solitary-wave solutions in the form of a spatially-localized wave that travels through the system without decaying or changing form. Solitary-wave solutions occur in many well-known particle-chain systems, such as the Toda lattice \cite{Faddeev1, Toda1}.

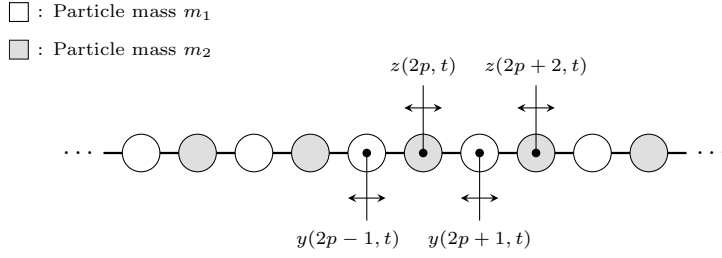
\begin{figure}
\centering
\begin{tikzpicture}
[xscale=1,>=stealth,yscale=1]

\draw [line width=0.3mm] (5.5,0) node[left] {$\ldots$} -- (13.25,0) node[right] {$\ldots$};
\fill [white] (6,0) circle (0.25);
\fill [white] (6.75,0) circle (0.25);
\fill [gray,opacity=0.25] (6.75,0) circle (0.25);
\fill [white] (7.5,0) circle (0.25);
\fill [white] (8.25,0) circle (0.25);
\fill [gray,opacity=0.25] (8.25,0) circle (0.25);
\fill [white] (9,0) circle (0.25);
\fill [white] (9.75,0) circle (0.25);
\fill [gray,opacity=0.25] (9.75,0) circle (0.25);
\fill [white] (10.5,0) circle (0.25);
\fill [white] (11.25,0) circle (0.25);
\fill [gray,opacity=0.25] (11.25,0) circle (0.25);
\fill [white] (12,0) circle (0.25);
\fill [white] (12.75,0) circle (0.25);
\fill [gray,opacity=0.25] (12.75,0) circle (0.25);

\draw (6,0) circle (0.25);
\draw (6.75,0) circle (0.25);
\draw (7.5,0) circle (0.25);
\draw (8.25,0) circle (0.25);
\draw (9,0) circle (0.25);
\draw (9.75,0) circle (0.25);
\draw (10.5,0) circle (0.25);
\draw (11.25,0) circle (0.25);
\draw (12,0) circle (0.25);
\draw (12.75,0) circle (0.25);

\draw (4.25,2-0.5) -- (4.5,2-0.5) -- (4.5,1.875-0.5) node [right] {\scriptsize{: Particle mass $m_2$}} -- (4.5,1.75-0.5) -- (4.25,1.75-0.5) -- cycle;
\fill [gray,opacity=0.25] (4.25,2-0.5) -- (4.5,2-0.5) -- (4.5,1.75-0.5) -- (4.25,1.75-0.5) -- cycle;
\draw (4.25,2.25-0.5) -- (4.5,2.25-0.5) -- (4.5,2.375-0.5) node [right] {\scriptsize{: Particle mass $m_1$}} -- (4.5,2.5-0.5) -- (4.25,2.5-0.5) -- cycle;
\draw [white] (14.5,1.5) -- (14.25,1.5) -- (14.25,1.4) -- cycle;

\draw (9,0) -- (9,-0.9);
\filldraw[black] (9,0) circle (0.05);
\draw [<->] (8.75,-0.6) -- (9.25,-0.6);
\node at (8.75,-0.9) [below] {\scriptsize{$y(2p-1,t)$}};

\draw (9.75,0) -- (9.75,0.9);
\filldraw[black] (9.75,0) circle (0.05);
\draw [<->] (9.5,0.6) -- (10,0.6);
\node at (9.75,0.9) [above] {\scriptsize{$z(2p,t)$}};

\draw (10.5,0) -- (10.5,-0.9);
\filldraw[black] (10.5,0) circle (0.05);
\draw [<->] (10.25,-0.6) -- (10.75,-0.6);
\node at (10.5,-0.9) [below] {\scriptsize{$y(2p+1,t)$}};

\draw (11.25,0) -- (11.25,0.9);
\filldraw[black] (11.25,0) circle (0.05);
\draw [<->] (11,0.6) -- (11.5,0.6);
\node at (11.25,0.9) [above] {\scriptsize{$z(2p+2,t)$}};

\end{tikzpicture}
\caption{Setup for the diatomic Toda chain with alternating particle masses. Instead of individual particle labels, we label each pair. We use $y(2p-1,t)$ to denote the displacement of the heavy particles, and we use $z(2p,t)$ to denote the displacement of the light particles.
}\label{F:AlternatingPic}
\end{figure}

In the present paper, we study periodic heterogenous chains in which the component particles have different masses. Specifically, as we illustrate with a schematic in Figure \ref{F:IntroPic}(b), we examine period-2 particle chains in which particles of different masses alternate with each other. We suppose that the mass ratio $m_2/m_1$ is small, so that $0 < m_2/m_1 \ll 1$. We thus say that particle one is the `heavy' particle and particle two is the `light' particle. Imposing this configuration and scaling the system \eqref{0:G} yields
\begin{align}
	\label{0:2P1}\ddot{y}(2p-1,t) &= \phi'\left[y(2p-1,t) - z(2p-2,t)\right]- \phi'\left[z(2p,t) - y(2p-1,t)\right] \,,\\
	\label{0:2P2}\eps\ddot{z}(2p,t) &= \phi'\left[z(2p,t) - y(2p-1,t)\right]-\phi'\left[y(2p+1,t) - z(2p,t)\right] \,,
\end{align}
where $y(2p-1,t)$ and $z(2p)$, respectively, are the displacements of the heavy and light particles, and the mass ratio is $m_2/m_1 = \eps$. We illustrate this configuration in Figure \ref{F:AlternatingPic}. We see that $y(2p-1,t)$ is defined only for odd particles, and $z(2p,t)$ is defined only for even particles.

To fully specify the solution to the system (\ref{0:2P1},\ref{0:2P2}), we also require initial conditions for $y(2p-1,t)$, $z(2p,t)$, and their first-order time derivatives. See Section \ref{S:2.1} for a discussion of these initial conditions. Finally, we specify the spatial boundary conditions as $n \rightarrow \inf$ so that they are consistent with the initial conditions in the same limit.

To find the leading-order behavior of (\ref{0:2P1},\ref{0:2P2}), we set $\eps = 0$ in \eqref{0:2P2} to get
\begin{equation}
	z(2p,t) = \tfrac{1}{2}\left[y(2p+1,t) + y(2p-1,t)\right]\label{0:zgen}
\end{equation}
We can therefore eliminate $z(n,t)$ from equation \eqref{0:2P1} to obtain
\begin{equation}
	\ddot{y}(2p-1,t) =  \phi'\left[\tfrac{1}{2}(y(2p-1,t) - y(2p-3,t))\right]-\phi'\left[\tfrac{1}{2}(y(2p+1,t) - y(2p-1,t))\right],\label{0:ygen}
\end{equation}
which is a scaled version of the nearest-neighbor equation \eqref{0:G}. Consequently, one can obtain solutions to the $\eps = 0$ version of the system (\ref{0:2P1},\ref{0:2P2}) by appropriately scaling solutions to the original nearest-neighbor equation. We therefore use solitary-wave solutions to the $\eps = 0$ case as the basis for an asymptotic study of the full system (\ref{0:2P1},\ref{0:2P2}), in which $\eps$ is a small positive parameter.

We will use techniques from exponential asymptotics to show that the period-2 Toda lattice, which is of the form (\ref{0:2P1},\ref{0:2P2}), permits nanopteron solutions, and we will determine the functional form of these solutions. 


\subsection{Formulation and Series Solution}\label{S:2.1}

We consider the period-2 form of the Toda-lattice equation associated with the system (\ref{0:2P1},\ref{0:2P2}), written as
\begin{align}
	\ddot{y}(2p-1,t) &=  \e^{-[y(2p-1,t)-z(2p-2,t)]}-\e^{-[z(2p,t)-y(2p-1,t)]} \,,\label{2.1 Toda1}\\
	\eps\ddot{z}(2p,t) &=  \e^{-[z(2p,t)-y(2p-1,t)]} -\e^{-[y(2p+1,t)-z(2p,t)]}\,,\label{2.1 Toda2}
\end{align}
where $0 < \eps \ll 1$, and we recall that $\eps$ is proportional to the ratio between the masses of the light and heavy particles in a chain. In general, the system (\ref{2.1 Toda1},\ref{2.1 Toda2}) requires two initial conditions each at $t=0$ for $y(2p-1,t)$ and $z(2p,t)$. However, to construct a nonlocal solitary-wave solution, we instead begin with a solution to the leading-order system (associated with $\eps = 0$), which takes the form given in (\ref{0:zgen},\ref{0:ygen}). In particular, \eqref{0:ygen} is a monoatomic Toda-lattice equation, for which we can obtain exact soliton solutions (given explicitly in \eqref{0:y0new}). We denote these leading-order solutions by $y_0(2p-1,t)$ and $z_0(2p,t)$. In practice, this is equivalent to posing the problem with initial conditions 
\begin{equation}
	y(n,0) = y_0(2p-1,0)\,,\enskip \dot{y}(2p-1,0) = \dot{y}_0(n,0)\,, \enskip z(2p,0) = z_0(2p,0)\,,\enskip \dot{z}(n,0) = \dot{z}_0(2p,0)\,.
\end{equation}
Finally, we specify that the particles ahead of the solitary-wave front (which travels from left to right) are undisturbed. This necessitates that
\begin{equation}
	y(2p-1,t) \rightarrow y_0(2p-1,t) \quad \mathrm{as} \quad p \rightarrow \inf\,,\qquad z(2p,t) \rightarrow z_0(2p,t) \quad \mathrm{as} \quad p \rightarrow \inf\,.
\end{equation}
This ensures that nonlocal oscillations remain in the wake of the leading-order soliton and consequently that any ensuing nanopteron solutions are one-sided.

We expand the dependent variables as series in $\eps$ and write
\begin{equation}\label{2.1 series}
	y(2p-1,t) \sim y_0\sum_{j = 2}^{\inf}\eps^j y_j(2p-1,t)\,, \qquad z(2p,t) \sim \sum_{j=0}^{\inf} \eps^j z_j(2p,t)\,.
\end{equation}
Inserting equation \eqref{2.1 series} into equations (\ref{2.1 Toda1},\ref{2.1 Toda2}) yields
\begin{align}
\nonumber\sum_{j = 0}^{\inf}\eps^j \ddot{y}_j(2p-1,t) = \exp\Bigg(-\sum_{j = 0}^{\inf}&\eps^j [y_j(2p-1,t) - z_j(2p-2,t)]\Bigg) - \\
	&\exp \Bigg(-\sum_{j = 0}^{\inf}\eps^j [z_j(2p,t) - y_j(2p-1,t)]\Bigg)
\label{2.1 gov1}\,,\\
\nonumber\sum_{j = 0}^{\inf}\eps^{j+1} \ddot{z}_j(2p,t) = \exp\Bigg(-\sum_{j = 0}^{\inf}&\eps^j [z_j(2p,t) - y_j(2p-1,t)]\Bigg) \\
	&\exp\Bigg(-\sum_{j = 0}^{\inf}\eps^j [y_j(2p+1,t) - z_j(2p,t)]\Bigg)
\label{2.1 gov2}\,.
\end{align}
Using a standard asymptotic power-series approach to (\ref{2.1 gov1},\ref{2.1 gov2}) entails expanding about $\eps = 0$ and matching to leading order in $\eps$. The resulting equation for $y_0(2p-1,t)$ is of the form given in \eqref{0:ygen}, and that the equation for $z_0(2p,t)$ is given by \eqref{0:zgen}. The leading-order system has a soliton solution
\begin{align}\label{0:y0new}
	y_0(2p-1,t) &= 2\log\left[\frac{1+\exp(2\kappa(p-1)-t\sqrt{2}\sinh{\kappa})}{1+\exp(2\kappa p-t\sqrt{2}\sinh{\kappa})}\right]\\
	z_0(2p,t) & = \log\left[\frac{1+\exp(2\kappa(p-1)-t\sqrt{2}\sinh{\kappa})}{1+\exp(2\kappa (p+1)-t\sqrt{2}\sinh{\kappa})}\right]\,,\label{0:z0new}
\end{align}
where $\kappa$ is an arbitrary constant that affects both the speed and width of the leading-order soliton solution. The speed of the soliton is $\sqrt{2}\, \kappa/\sinh{\kappa}$, and the width is proportional to $1/\kappa$.

We are interested in traveling-wave solutions, so we define the moving frame
\begin{equation}
	\xi = t - \frac{\sqrt{2}\,\kappa \,p}{\sinh \kappa} = t - \lambda \,p\,,\label{2.1 xi}
\end{equation}
where $\xi$ moves with the leading-order soliton, which propagates from left to right in $p$ as $t$ increases. The region ahead of the wave front corresponds to $t < \lambda p$, and hence $\xi < 0$, whereas the region in the wake of the wave front corresponds to $t > \lambda p$, or $\xi > 0$. The leading-order equations become
\begin{align}
	\label{2.N y0}y_0(2p-1,t) &= y_0(\xi) = 2\log\left[\frac{1+\exp(-2\sinh\kappa\,\xi-2\kappa)}{1+\exp(-2\sinh\kappa\,\xi)}\right] \,,\\
	\label{2.N z0}z_0(2p,t) &= z_0(\xi)= \log\left[\frac{1+\exp(-2\sinh\kappa\,\xi-2\kappa)}{1+\exp(-2\sinh\kappa\,\xi+2\kappa)}\right]\,.
\end{align}

We thereby represent the Toda system by the coupled delay-differential equations (DDEs)
\begin{align}
	\label{2.1 Toda1b}	\sum_{j = 0}^{\inf}\eps^j y''_j(\xi) &= \exp\Bigg(-\sum_{j = 0}^{\inf}\eps^j [y_j(\xi,t) - z_j(\xi+\lambda)]\Bigg)-\exp \Bigg(-\sum_{j = 0}^{\inf}\eps^j [z_j(\xi) - y_j(\xi)]\Bigg) \,,\\
	\label{2.1 Toda2b}	\sum_{j = 0}^{\inf}\eps^{j+1} z''_j(\xi) &= \exp\Bigg(-\sum_{j = 0}^{\inf}\eps^j [z_j(\xi) - y_j(\xi)]\Bigg)-\exp\Bigg(-\sum_{j = 0}^{\inf}\eps^j [y_j(\xi-\lambda) - z_j(\xi)]\Bigg) \,,
\end{align}
where $'$ denotes differentiation with respect to $\xi$, and $\lambda$ is defined in \eqref{2.1 xi}.

Previous studies, such as \cite{Ahnert1}, examined systems of the form (\ref{2.1 Toda1},\ref{2.1 Toda2}) by writing the solution in terms of integral equations. In contrast, we apply asymptotic-series methods directly to the differential--difference system in (\ref{2.1 Toda1b},\ref{2.1 Toda2b}). By using these methods, we will be able to directly study exponentially small oscillations in the solutions.

One can compute subsequent terms in the series using equations (\ref{2.1 gov1},\ref{2.1 gov2}) and the recursion relation
\begin{align}
	\label{2.2 LOT1} y''_j(\xi) = -&[y_j(\xi) - z_j(\xi+\lambda)]\e^{-[y_0(\xi) -z_0(\xi+\lambda)]} + [z_j(\xi) - y_j(\xi)]\e^{-[z_0(\xi) -y_0(\xi)]} + \ldots\,,\\
	\label{2.2 LOT2} z''_{j-1}(\xi) = -&[z_j(\xi) - y_j(\xi)]\e^{-[z_0(\xi) -y_0(\xi)]}+ [y_j(\xi-\lambda) - z_j(\xi)]\e^{-[y_0(\xi-\lambda) -z_0(\xi)]} + \ldots\,,
\end{align} 
where the omitted terms are subdominant as $j$ becomes large. It typically is difficult to solve equations such as (\ref{2.2 LOT1},\ref{2.2 LOT2}) analytically in terms of a closed-form expression, although Vainchtein et al. \cite{Vainchtein1} derived an exact solution for the correction to the leading-order behavior for the diatomic Toda lattice in terms of hypergeometric functions. In contrast, the present method does not require the computation of these correction terms. We will compare our result to the exact calculation from \cite{Vainchtein1} in Section \ref{S:Results}.

The wave train has exponentially small amplitude, so it lies beyond the reach of an asymptotic power-series approach. Consequently, it will not be helpful to obtain correction terms to the asymptotic power series. Instead, we will apply exponential asymptotic methods to determine the dynamics of these trailing waves.


\subsection{Late-Order Terms}\label{S:2.2}

As we discussed in Section \ref{Method}, we need to ascertain the behavior of the terms in the series (\ref{2.1 series}) in the limit that $j \rightarrow \inf$ to determine the exponentially small component of the solution. Unfortunately, it is difficult to exactly compute even the first correction term of the series.  We note from (\ref{2.2 LOT1},\ref{2.2 LOT2}) that obtaining the behavior of $y_j$ and $z_j$ for large $j$ requires differentiating $z_{j-1}$ twice. This suggests that we will encounter repeated differentiations of the form discussed in Section \ref{Method}. Consequently, we follow the process described by Chapman et al. \cite{Chapman1} and apply a factorial-over-power ansatz similar to equation \eqref{ch1:ansatz} to approximate these late-order terms in the $j\rightarrow \inf$ limit.

We must proceed with care, as (\ref{2.2 LOT1},\ref{2.2 LOT2}) are differential--difference equations; however, the method devised by Chapman et al. \cite{Chapman1} was developed for differential equations. Fortunately, in our subsequent analysis, we will see that the difference terms decouple at relevant orders of $j$ in the limit that $j \rightarrow \inf$, and that the method of \cite{Chapman1} may be applied in straightforward fashion to the expression that remains.

Hence, we assume that the late-order terms are given by a sum of expressions of the form
\begin{equation}\label{2.2 ansatz}
	y_r(\xi) \sim \frac{Y(\xi) \Gamma(2r + \gamma_1)}{\chi(\xi)^{2r + \gamma_1}}\,, \qquad z_j(\xi) \sim \frac{Z(\xi) \Gamma(2r + \gamma_2)}{\chi(\xi)^{2r + \gamma_2}}\,,
\end{equation}
where $\gamma_1$ and $\gamma_2$ are constants, and $Y$, $Z$, and $\chi$ are functions of $\xi$. We also analytically continue $\xi$ so that $\xi \in \mathbb{C}$. Recall that the singulant $\chi$ is particularly important, as it permits singularities in the leading-order behavior to propagate into late-order terms. It is equal to $0$ at points $\xi = \xi_s$ in the complex plane at which the leading-order behavior is singular. We observed that obtaining $z_j$ required two differentiations of $z_{j-1}$, so we expect the strength of the singularity to increase by 2 between $r = j-1$ and $r = j$. As each successive term in the sequence requires second-order differentiation, we also expect the argument of the gamma function to increase by 2 in each successive term. This accounts for the form in the ansatz (\ref{2.2 ansatz}). 

As we noted above, the late-order terms are given by the sum of expressions of the form \eqref{2.2 ansatz}. In this case, each expression is associated with a different singularity in the analytic continuation of $z_0$. We denote the locations of these singularities by $\xi = \xi_{s}$ (where $s \in \{1, \ldots ,4\}$). They occur at
\begin{align}
	\xi_{1,\pm} = \frac{2\kappa \pm i\pi}{\sqrt{2}\sinh{\kappa}}\,,\qquad \xi_{2,\pm} = \frac{-2\kappa \pm i\pi}{\sqrt{2}\sinh{\kappa}}\,.\label{2.2 xis}
\end{align}
The analytic continuation of $z_0$ contains other singularities; however, late-order term asymptotic behavior at any point is exponentially dominated by the nearest singularities to that point \cite{Dingle1}. Consequently, we can neglect singularities that are farther away from the real axis. 

The leading-order behavior $z_0$ is singular at the points in \eqref{2.2 xis}. Hence, each of these points is associated with a late-order ansatz of the form in \eqref{2.2 ansatz}, with $\chi = 0$ at the corresponding singular point $\xi_s$.

We note that this configuration of analytically-continued singularities, in which one or more singularities extend in the positive and negative imaginary directions from the traveling-wave front, appears often in the study of nanoptera and related problems. It is common for leading-order traveling-wave solutions to contain trains of singularities, which extend in both imaginary directions, that determine the sizes and locations of exponentially small oscillations in the associated singularly-perturbed problem. This occurs, for example, in the study of the singularly-perturbed fifth-order Korteweg--de Vries equation \cite{Grimshaw1,Trinh5}.

Our leading-order solution \eqref{0:y0new}--\eqref{0:z0new} contains trains of singularity pairs with the same real part. Such behavior was also observed in \cite{Gelfreich1} and \cite{Chapman11} in exponential asymptotic analysis of homoclinic orbits and snaking phenomena, respectively. In particular, singularity pairs like those described in \eqref{2.2 xis} raise the possibility of solutions in which oscillatory terms cancel precisely. 

These pairs play an important role in \cite{Alfimov1, Calvo1, Lustri1}, with each singularity producing oscillatory contributions of identical amplitude. For particular choices of a small parameter that satisfies an orthogonality condition, each of these systems can be tuned to produce solutions in which any oscillations cancel exactly. 

We will identify a similar orthogonality condition for the 2-periodic Toda lattice using a direct Stokes-curve analysis. The presence of multiple singularities allows contributions from multiple Stokes curves, and we will see that the orthogonality condition describes solutions in which the oscillatory contributions cancel precisely.


\subsubsection{Calculating $\chi(\xi)$}

Applying the late-order ansatz (\ref{2.2 ansatz}) to the first governing equation (\ref{2.2 LOT1}) shows that we need $\gamma_1 + 2 = \gamma_2$ to obtain a nontrivial balance. This implies that $y_r = \mathcal{O}(z_{r-1})$ and hence that $z_r$ dominates $y_r$ in the late-order recursion equations (\ref{2.2 LOT1},\ref{2.2 LOT2}).

Applying the late-order ansatz (\ref{2.2 ansatz}) to the second governing equation (\ref{2.2 LOT2}) subsequently gives
\begin{align}
	\frac{(\chi'(\xi))^2 Z(\xi) \Gamma(2j + \gamma_2)}{\chi(\xi)^{2j+\gamma_2}} - \frac{2\chi'(\xi) Z'(\xi) \Gamma(2j + \gamma_2-1)}{\chi(\xi)^{2j+\gamma_2-1}}&\nonumber \\
	-\frac{\chi''(\xi) Z(\xi) \Gamma(2j + \gamma_2-1)}{\chi(\xi)^{2j+\gamma_2-1}} = -2\e^{-\tfrac{1}{2}(y_0(\xi-\lambda)-y_0(\xi))}&\frac{ Z(\xi) \Gamma(2j + \gamma_2)}{\chi(\xi)^{2j+\gamma_2}} +\ldots\,,\label{A.2 LOEQ}
\end{align}
where the omitted terms are no larger that $\mathcal{O}(z_{r-1})$ in the $r \rightarrow \inf$ limit. We now equate terms that are $\mathcal{O}(z_r)$ as $r \rightarrow \inf$ to obtain the singulant equation
\begin{equation}\label{A.2 singulant1}
	\left[\chi'(\xi)\right]^2 = -2\e^{-\tfrac{1}{2}[y_0(\xi-\lambda)-y_0(\xi)]}\,, \qquad \chi = 0 \quad \mathrm{at} \quad \xi = \xi_s\,.
\end{equation}
We can write the solution to this expression as
\begin{equation}\label{A.2 singulant}
	\chi(\xi) = \pm i \sqrt{2}\int_{\xi_s}^{\xi} \sqrt{\frac{(1+\exp(-\sqrt{2}\sinh{\kappa}\, s + 2\kappa))(1+\exp(-\sqrt{2}\sinh{\kappa}\, s - 2\kappa))}{(1+\exp(-\sqrt{2}\sinh{\kappa}\, s))^2}}\,\mathrm{d} s\,.
	\end{equation}
Evaluating this integral with the singularity at $\xi = \xi_{1,+}$ gives
\begin{align}
	\nonumber\chi_{1,+} = \pm \Bigg( i \xi \sqrt{2} -i \pi + \frac{\pi}{\sinh{\kappa}}+ 2\log\left[\frac{i (\e^{\nu}-1)\sinh{\kappa} + 2\e^{\nu/2}\sqrt{\cosh(\kappa-\nu/2)\cosh(\kappa+\nu/2)}}{(\e^{\nu}+1)\cosh{\kappa}}\right]&\\
+ \frac{i}{\sinh{\kappa}} \log\left[\frac{\e^{\nu} + \cosh(2\kappa) +2\e^{\nu/2}\sqrt{\cosh(\kappa-\nu/2)\cosh(\kappa+\nu/2)}}{1 + \e^{\nu}\cosh(2\kappa) + 2\e^{\nu/2}\sqrt{\cosh(\kappa-\nu/2)\cosh(\kappa+\nu/2)}}\right]\Bigg)&\,,\label{singulantexact}
\end{align}
where $\nu = \sqrt{2}\sinh{\kappa} \,\xi$.

\begin{figure}
\centering
\subfloat[$\mathrm{Re}(\chi)$]{
\begin{tikzpicture}
[xscale=0.5,>=stealth,yscale=0.15]

\draw[thick] (-5,2.67) -- (5,2.67);
\node at (0,2.67) [above left] {\scriptsize{$2.67$}};

\node at (5,2.67) [above left] {\scriptsize{$\chi_{1,\pm}\quad\chi_{2,\pm}$}};

\draw[->] (-5.5,0) -- (5.5,0) node[right] {\scriptsize{$\xi$}};
\draw[->] (0,-13) -- (0,13) node[above] {\scriptsize{$\mathrm{Re}(\chi)$}};

\draw (-4,0.5) -- (-4,-0.5) node[below] {\scriptsize{$-4$}};
\draw (4,0.5) -- (4,-0.5) node[below] {\scriptsize{$4$}};
\draw (0.15,10) -- (-0.15,10) node[left] {\scriptsize{$10$}};
\draw (0.15,-10) -- (-0.15,-10) node[left] {\scriptsize{$-10$}};

\draw[white] (-7,-16) -- (7,-16) -- (7,16) -- (-7,16);

\end{tikzpicture}
}
\subfloat[$\mathrm{Im}(\chi)$]{
\begin{tikzpicture}
[xscale=0.5,>=stealth,yscale=0.15]

\draw[thick] plot [smooth] file {chi1_imag.txt} node[right] {\scriptsize{$\chi_{1,+}$}};
\draw[thick] plot [smooth] file {chi2_imag.txt} node[right] {\scriptsize{$\chi_{1,-}$}};
\draw[thick] plot [smooth] file {chi3_imag.txt};
\draw[thick] plot [smooth] file {chi4_imag.txt};
\node at (5,11.2) [right] {\scriptsize{$\chi_{2,+}$}};
\node at (5,-11.2) [right] {\scriptsize{$\chi_{2,-}$}};

\draw[->] (-5.5,0) -- (5.5,0) node[right] {\scriptsize{$\xi$}};
\draw[->] (0,-13) -- (0,13) node[above] {\scriptsize{$\mathrm{Im}(\chi)$}};

\draw (-4,0.5) -- (-4,-0.5) node[below] {\scriptsize{$-4$}};
\draw (4,0.5) -- (4,-0.5) node[below] {\scriptsize{$4$}};

\draw (-4,0.5) -- (-4,-0.5) node[below] {\scriptsize{$-4$}};
\draw (4,0.5) -- (4,-0.5) node[below] {\scriptsize{$4$}};
\draw (0.15,10) -- (-0.15,10) node[left] {\scriptsize{$10$}};
\draw (0.15,-10) -- (-0.15,-10) node[left] {\scriptsize{$-10$}};

\draw[white] (-7,-16) -- (7,-16) -- (7,16) -- (-7,16);

\end{tikzpicture}
}

\caption{The (a) real part of $\chi_{1,\pm}$ and $\chi_{2,\pm}$ and (b) the imaginary part of these functions. All four singulants have the same constant positive real part. Each singulant has $0$ imaginary part at one real value of $\chi$. This is the location of the associated Stokes curve. We illustrate the full Stokes structure for $\xi \in \mathbb{C}$ in Figure \ref{F:StokesStruct}. We also see that $\mathrm{Im}(\chi)$ tends to a straight line away from the origin. In Section \ref{S:AsymptoticChi}, we will show that $\mathrm{Im}(\chi'(\xi)) \rightarrow \pm\sqrt{2}$ as $\xi \rightarrow \pm\inf$.}\label{F:chi}
\end{figure}
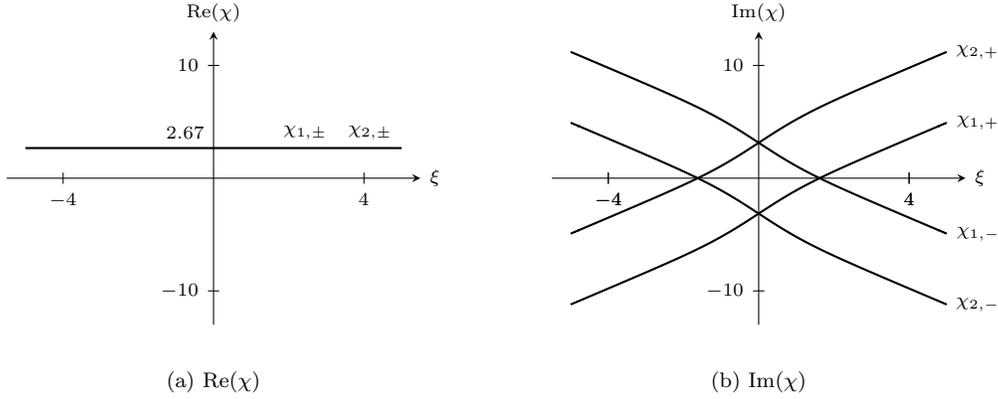

The expression for $\xi_{1,-}$ is given by the complex conjugate of \eqref{singulantexact}, and $\xi_{2,\pm}$ produces an expression that is the same except for a shift in $\xi$ by $2\sqrt{2}/\sinh{\kappa}$ in the negative real direction. In each case, the real part of the singulant is constant. We show these functions in Figure \ref{F:chi}.

From \eqref{A.2 singulant}, we know that there are two possible singulants (which differ only in sign) that are associated with each singularity: one has a positive component on the real $\xi$-axis, and the other has a negative real part. Because Stokes switching can only occur when the real part of the singulant is positive, we can ignore the singulants with negative real part, as these singulants do not contribute to the Stokes-switching behavior. This corresponds to choosing the positive sign in \eqref{A.2 singulant} and hence in \eqref{singulantexact}.

Finally, we can also evaluate the integral between two singular points directly to obtain 
\begin{equation}\label{sing2pii}
	i \sqrt{2}\int_{\xi_{2,+}}^{\xi_{1,+}} \sqrt{\frac{(1+\exp(-\sqrt{2}\sinh{\kappa}\, s + 2\kappa))(1+\exp(-\sqrt{2}\sinh{\kappa}\, s - 2\kappa))}{(1+\exp(-\sqrt{2}\sinh{\kappa}\, s))^2}}\,\mathrm{d} s = 2 \pi i\,.
\end{equation}
This result will be useful in subsequent analysis.


\subsubsection{Asymptotic Behavior of  $\chi(\xi)$}\label{S:AsymptoticChi}

It is also useful to know the behavior of the singulant in both the neighborhood of the singularity and far from the wave front (i.e., as $\xi \rightarrow \pm\inf$, which is the `far field'). We thus examine the late-order terms associated with $\xi_{1,\pm}$, and we then state the corresponding results for the terms associated with $\xi_{2,\pm}$.

As $\xi \rightarrow \pm \inf$, it is clear that $y_0(\xi-\lambda)-y_0(\xi) \rightarrow 0$ and therefore thus $\chi'(\xi) \sim \pm i \sqrt{2}$. 

In an inner region near the singularity at $\xi = \xi_{1,\pm}$, we see that
\begin{align}
	z_0(\xi) &\sim -\log(\xi-\xi_{1,\pm}) + \log(1-\e^{-4\kappa}) - \log(\sqrt{2}\sinh\kappa) + \mathcal{O}(\xi-\xi_{1,\pm})\,,\label{z0inner}\\
	y_0(\xi) &\sim -2\log(\xi-\xi_{1,\pm})  + 2\log(1-\e^{-2\kappa})  - 2\log(\sqrt{2}\sinh\kappa) + \mathcal{O}(\xi-\xi_{1,\pm})\,,\label{y0inner}\\
	y_0(\xi-\lambda) &\sim 2\log(1+\e^{-2\kappa}) + \mathcal{O}(\xi-\xi_{1,\pm})\,.\label{y01inner}
\end{align}

In the inner region, we find that $\chi'(\xi) \sim \beta (\xi-\xi_{1,\pm})^{1/2}$ as $\xi \rightarrow \xi_{1,\pm}$, where $\beta = 2^{1/4} i \sqrt{\cosh \kappa}$. Therefore, the local behavior of the singulant is 
\begin{equation}\label{2.N SingulantLocal}
	\chi \sim \tfrac{2\beta}{3}(\xi-\xi_{1,\pm})^{3/2} \qquad \mathrm{as} \qquad \xi \rightarrow \xi_{1,\pm}
\end{equation}
in the neighborhood of the singularity. By performing a similar local analysis near $\xi_{2,\pm}$, we find that \eqref{2.N SingulantLocal} also holds for both $\xi_s = \xi_{1,\pm}$ and $\xi_s = \xi_{2,\pm}$.


\subsubsection{Calculating $Z(\xi)$}

We now match equation \eqref{A.2 LOEQ} at the next order in $j$ by equating terms of $\mathcal{O}(z_{r-1/2})$ as $r \rightarrow \inf$. This yields the equation
\begin{equation}
	-\frac{2 Z'(\xi) \chi'(\xi) \Gamma(2j + \gamma_2-1)}{\chi(\xi)^{2j+\gamma_2-1}} + \frac{\chi''(\xi) Z(\xi) \Gamma(2j + \gamma_2-1)}{\chi(\xi)^{2j+\gamma_2-1}} = 0\,,
\end{equation}
which simplifies to the equation
\begin{equation}\label{A.2 prefactor1}
	2Z'(\xi)\chi'(\xi) + Z(\xi)\chi''(\xi) = 0
\end{equation}
 for the prefactor $Z(\xi)$. We then integrate \eqref{A.2 prefactor1} to obtain
\begin{equation}\label{A.2 prefactor}
	Z(\xi) = \frac{\Lambda}{\sqrt{\chi'(\xi)}}\,,
\end{equation}
where $\Lambda$ is a constant that one can determine by considering the behavior of the solution near $\xi = \xi_s$ and comparing it to the leading-order behavior in this local neighborhood. We perform this local analysis in detail in Appendix \ref{S:2.2.5.2} and obtain
\begin{align}
	\frac{\Lambda_{1,+}}{({\cosh{\kappa}})^{1/6}}&\approx 0.545 + 0.314i\,,\qquad& \frac{\Lambda_{1,-}}{({\cosh{\kappa}})^{1/6}}&\approx 0.545 - 0.314i\,,\\
\qquad \frac{\Lambda_{2,+}}{({\cosh{\kappa}})^{1/6}}&\approx -0.545 - 0.314i\,,\qquad &\frac{\Lambda_{2,-}}{({\cosh{\kappa}})^{1/6}}&\approx -0.545 + 0.314i\,.
\end{align}
Recalling \eqref{2.N SingulantLocal}, we see that 
\begin{equation}\label{2.N PrefactorLocal}
	Z_{s,\pm}(\xi) \sim \frac{\Lambda_{s,\pm}}{\sqrt{\beta(\xi-\xi_{s,\pm})^{1/2}}} \qquad \mathrm{as} \qquad \xi \rightarrow \xi_{s,\pm}\,.
\end{equation}


\subsubsection{Calculating $\gamma_2$}

To determine $\gamma_2$, we need to consider the behavior of the solution near the singularity under consideration, where the asymptotic series (\ref{2.1 series}) breaks down. To determine the correct scaling, we examine the local behavior of $y_0$ and $z_0$ and hence of $\chi$.

The local expression for the singulant in the neighborhood of the singularity is given in \eqref{2.N SingulantLocal}, and the local expression for the prefactor is given in \eqref{2.N PrefactorLocal}. Consequently, the local behavior of the factorial-over-power ansatz (\ref{2.2 ansatz}) in the neighborhood of a singularity at $\xi = \xi_s$ is 
\begin{equation}\label{2.2.5 ansatz}
	z_j(n,t) \sim \frac{\Lambda\Gamma(2j+\gamma_2)}{\sqrt{\beta(\xi-\xi_s)^{1/2}}\left[\tfrac{2\beta}{3}(\xi-\xi_s)^{3/2}\right]^{2j+\gamma_2}}
\qquad \mathrm{as} \qquad \xi	\rightarrow \xi_s \quad \mathrm{and} \quad j\rightarrow \inf\,.
\end{equation}
For \eqref{2.2.5 ansatz} to be consistent with the leading-order behavior \eqref{2.N z0}, the singularity in the late-order terms as $j \rightarrow 0$ must have the same strength as the singularity in the leading-order behavior.
Because the leading-order behavior has a logarithmic singularity, we require that the strength of the singularity in the late-order terms tend to $0$ when $j = 0$. It then follows that $1/4 + 2\gamma_2/3 = 0$, which gives $\gamma_2 = -1/6$. 

Therefore, as $j \rightarrow \inf$, one can express the terms in the series expression \eqref{2.1 series} as
\begin{align}
	z_j(\xi) \sim \frac{\Lambda_{1,+}\Gamma(2j - 1/6)}{\sqrt{\chi_{1,+}'(\xi)}[\chi_{1,+}(\xi)]^{2j - 1/6}} + \frac{\Lambda_{1,-}\Gamma(2j - 1/6)}{\sqrt{\chi_{1,-}'(\xi)}[\chi_{1,+}(\xi)]^{2j - 1/6}}&\nonumber\\
	+ \frac{\Lambda_{2,+}\Gamma(2j - 1/6)}{\sqrt{\chi_{2,+}'(\xi)}[\chi_{2,+}(\xi)]^{2j - 1/6}}+
	&\frac{\Lambda_{2,-}\Gamma(2j - 1/6)}{\sqrt{\chi_{2,-}'(\xi)}[\chi_{2,-}(\xi)]^{2j - 1/6}}\,,\label{2.2 zjlate}
\end{align}
where $\chi$ is given in \eqref{singulantexact} and $Z$ is given in \eqref{A.2 prefactor}.


\subsection{Exponential Asymptotics}\label{S:2.3}

The key idea of exponential asymptotics is that one can truncate an asymptotic series optimally at some term number $N_{\mathrm{opt}}$ such that the remainder is exponentially small. The solution can thus be expressed as
\begin{equation}\label{2.2 seriesopt}
	y(\xi) \sim \sum_{r = 0}^{N_{\mathrm{opt}}}\eps^r y_r(\xi) + y_{\mathrm{exp}}(\xi)\,, \qquad z(\xi) \sim \sum_{r=0}^{N_{\mathrm{opt}}} \eps^r z_r(\xi) + z_{\mathrm{exp}}(\xi)\,,
\end{equation}
where $y_{\mathrm{exp}}$ and $z_{\mathrm{exp}}$ are exponentially small as $\eps \rightarrow 0$.

Recall from Section \ref{Method} that one can determine the exponentially small contribution to the asymptotic solution from the form of the late-order terms, and that this contribution switches rapidly across Stokes curves. We will apply exponential asymptotic methods to determine the form of the late-order terms and the associated Stokes-curve locations, and we will thereby determine the form of the oscillatory behavior that changes as the Stokes curves are crossed.


\subsubsection{Stokes Structure}\label{S:2.3A}

Recall that Stokes curves are curves in the complex plane that satisfy $\mathrm{Im}(\chi) = 0$ and $\mathrm{Re}(\chi) > 0$. We thus see that $z_{\mathrm{exp}}$ is exponentially small as $\eps \rightarrow 0$, as both $\mathrm{Re}(\chi_{1,\pm}) > 0$ and $\mathrm{Re}(\chi_{2,\pm}) > 0$. 

\begin{figure}
\centering
\begin{tikzpicture}
[xscale=1,>=stealth,yscale=0.75]

\fill[lightgray,opacity=0.2] (5.5,4) -- plot [smooth] file {AS1_Upper.txt} -- (-5.5,4) -- cycle;
\fill[lightgray,opacity=0.2] plot [smooth] file {AS2_Upper.txt}  -- cycle;
\fill[lightgray,opacity=0.2] plot [smooth] file {AS2_Lower.txt}  -- cycle;
\fill[lightgray,opacity=0.2] (5.5,-4) -- plot [smooth] file {AS1_Lower.txt} -- (-5.5,-4) -- cycle;

\draw[thick] plot [smooth] file {AS1_Upper.txt};
\draw[thick] plot [smooth] file {AS2_Upper.txt};
\draw[thick,dashed] plot [smooth] file {SL1_Upper.txt};
\draw[thick,dashed] plot [smooth] file {SL2_Upper.txt};
\draw[thick,dashed] plot [smooth] file {SL3_Upper.txt};
\draw[thick,dashed] plot [smooth] file {SL4_Upper.txt};

\draw[thick] plot [smooth] file {AS1_Lower.txt};
\draw[thick] plot [smooth] file {AS2_Lower.txt};
\draw[thick,dashed] plot [smooth] file {SL1_Lower.txt};
\draw[thick,dashed] plot [smooth] file {SL2_Lower.txt};
\draw[thick,dashed] plot [smooth] file {SL3_Lower.txt};
\draw[thick,dashed] plot [smooth] file {SL4_Lower.txt};

\fill[white] (1.15,1.86) -- (1.15,1.92) -- (-1.15,1.92) -- (-1.15,1.86) -- cycle;
\fill[white] (1.15,-1.86) -- (1.15,-1.92) -- (-1.15,-1.92) -- (-1.15,-1.86) -- cycle;
\draw[decoration = {zigzag,segment length = 2mm, amplitude = 0.5mm},decorate, line width=0.25mm,black] (1.15,1.89)--(-1.15,1.89);
\draw[decoration = {zigzag,segment length = 2mm, amplitude = 0.5mm},decorate, line width=0.25mm,black] (1.15,-1.89)--(-1.15,-1.89);

\filldraw (1.2,1.89) ellipse (0.1 and 0.133);
\filldraw (-1.2,1.89) ellipse (0.1 and 0.133);
\filldraw (1.2,-1.89) ellipse (0.1 and 0.133);
\filldraw (-1.2,-1.89) ellipse (0.1 and 0.133);

\node at (1.4,1.9) [above right] {$\xi_{1,+}$};
\node at (1.4,-1.9) [below right] {$\xi_{1,-}$};
\node at (-1.4,1.9) [above left] {$\xi_{2,+}$};
\node at (-1.4,-1.9) [below left] {$\xi_{2,-}$};

\draw[->] (0,-4.5) -- (0,4.5) node[above] {\scriptsize{$\mathrm{Im}(\xi)$}};
\draw[->] (-5.5,0) -- (5.5,0) node[right] {\scriptsize{$\mathrm{Re}(\xi)$}};

\node at (-4,1.85) [below] {\scriptsize{Anti-Stokes Curve: $\chi_{1,2}$}};
\node at (4,1.85) [below] {\scriptsize{Anti-Stokes Curve: $\chi_{1,2}$}};
\node at (-1.7,0.5) [left] {\scriptsize{Stokes Curve: $\chi_{2}$}};
\node at (1.7,0.5) [right] {\scriptsize{Stokes Curve: $\chi_{1}$}};

\end{tikzpicture}

\caption{The Stokes structure for $z(\xi)$. We represent Stokes curves as dashed curves and anti-Stokes curves as solid curves. 
The Stokes curves originate at the singularities (represented by black disks) in $z_0(\xi)$. Each of the anti-Stokes curves is associated with all four contributions. The shaded regions indicate regions in which $\mathrm{Re}{(\chi)} < 0$, so $z_{\mathrm{exp}}$ is exponentially large. Stokes switching requires $\mathrm{Re}{(\chi)} > 0$, so it cannot occur across any Stokes curve in these regions. The hatched lines represent branch cuts originating from the logarithmic singularities at $\xi_{1,\pm}$ and $\xi_{2,\pm}$. The Stokes curves that originate at $\xi_{1,\pm}$ and $\xi_{2,\pm}$ cause Stokes switching to occur in the exponentially small contributions associated with $\chi_{1,\pm}$ and $\chi_{2,\pm}$, respectively.  The region of $\xi \in \mathbb{R}$ that corresponds to the wake of the traveling wave front lies in the right-hand side of the figure ($\lambda p < t$), and the left-hand side corresponds to the undisturbed region ahead of the wave ($\lambda p >  t$). We expect each Stokes contribution to be absent on the left-hand side of the associated Stokes curve and switched on as the curve is crossed from left to right along the real axis. Recall that $\xi = t - \lambda p$, so $\xi \in \mathbb{R}$ is the region in which the quantities $p$ and $t$ take real, physically-valid values. 
}\label{F:StokesStruct}
\end{figure}
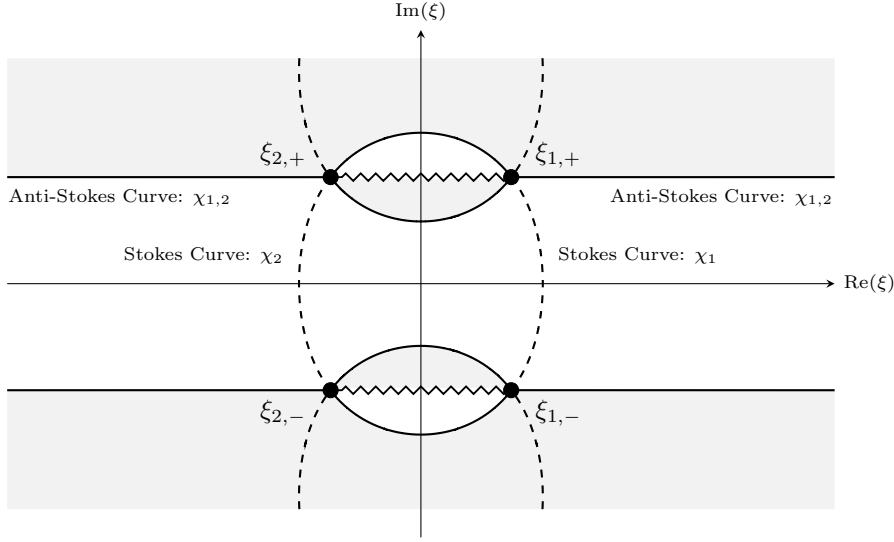

In Figure \ref{F:chi}, we show the imaginary parts of $\chi_{1,\pm}$ and $\chi_{2,\pm}$ with $\kappa = 1$. For each singulant, there is a single point that satisfies the Stokes-curve criteria on the real $\xi$ axis. In Figure \ref{F:StokesStruct}, we give a schematic to illustrate the behavior of the Stokes curves in the analytically-continued plane. The Stokes curves originate at the singularities in $\xi$, and we note that each of these singularities intersects the real $\xi$ axis at a single point.

We assume that the behavior that precedes the wave front is undisturbed, so we conclude that $z_{\mathrm{exp}}$ is present only to the right of these Stokes curves (i.e., in the wake of the leading-order soliton). As we noted in Section \ref{nano} and illustrated in Figure \ref{F:nanoptera}, more general nanopteron behavior can arise by relaxing this assumption, and that is one possible way to obtain two-sided nanoptera. However, modeling a wave front moving through an undisturbed chain necessitates that the region preceding the wave front does not include oscillatory behavior.


\subsubsection{Stokes Smoothing}\label{S:2.3B}

Now that we have obtained late-order terms, we truncate the series after $N$ terms to give
\begin{equation}\label{2.2 series}
	y(\xi)= \sum_{j = 0}^{N-1}\eps^j y_j(\xi) + S_N(\xi)\,, \qquad z(\xi)= \sum_{j=0}^{N-1} \eps^j z_j(\xi) + R_N(\xi)\,.
\end{equation}
To apply the exponential asymptotic technique of \cite{Chapman1}, we need to optimally truncate the series \eqref{2.2 series}. 
In the $\eps \rightarrow 0$ limit, we need a large number of terms for an optimal truncation. Following \cite{Boyd1}, we determine the optimal truncation by finding the point at which consecutive terms in the series are equal in size. That is,
\begin{equation}
	\left|\frac{\eps^{N+1} z_{N+1}(\xi)}{\eps^{N} z_{N}(\xi)}\right| \sim 1 \qquad \mathrm{as} \qquad \eps \rightarrow 0\,.
\end{equation}
Using the general form of the ansatz (\ref{2.2 ansatz}), we obtain $N \sim |\chi|/2\sqrt{\eps}$. Consequently, we set the optimal truncation point to be
\begin{equation}\label{2.3 truncation}
	N = \frac{|\chi|}{2\sqrt{\eps}} + \omega\,,
\end{equation}
where we choose $\omega \in [0,1)$ in a way that ensures that $N$ is an integer. We see that this quantity depends on $\chi$, which in turn depends on the position $\xi$. However, this does not cause any difficulties with the exponential asymptotic methodology, as $N$ depends continuously on $\xi$ to leading order in $\eps$. In particular, this ensures that any $\omega$ dependence vanishes from \eqref{inner_approx_noomega} and hence from the Stokes-switching behavior. For more detailed discussion on this choice of $N$, see \cite{Daalhuis1}.

We now apply the truncated series expression (\ref{2.2 series}) to (\ref{2.1 Toda1},\ref{2.1 Toda2}). Using the relation \eqref{0:zgen}, we obtain as $\eps \rightarrow 0$ that
\begin{align}
	 S_N''(\xi)& \sim R_N(\xi+\lambda)\e^{-\frac{1}{2}[y_0(\xi)-y_0(\xi+\lambda)]} + R_N(\xi)\e^{-\frac{1}{2}[y_0(\xi+\lambda)-y_0(\xi)]} + \ldots\,,\label{2.3 NewToda1}\\
	\eps  R_N''(\xi)& + 2\e^{-\frac{1}{2}[y_0(\xi-\lambda)-y_0(\xi)]} R_N(\xi) \sim -  \eps^N z''_{N-1}(\xi) + \ldots\,,\label{2.3 NewToda2}
\end{align}
where the terms that we omitted are smaller in magnitude than those that we retained by a factor of $\eps$ or more in the limit that $\eps \rightarrow 0$. Throughout the remainder of this discussion, we consider asymptotic behavior in the $\eps \rightarrow 0$ asymptotic limit. 

Applying the expression \eqref{A.2 singulant1} for the singulant and the late-order term ansatz (noting again that $N \rightarrow \inf$ as $\eps \rightarrow 0$), we obtain
\begin{equation}\label{2.3 RN}
	\eps R''_N(\xi) - \chi'(\xi)^2 R_N(\xi) \sim -\frac{\eps^N \chi'(\xi)^2 Z(\xi) \Gamma(2N -1/6)}{\chi(\xi)^{2N-1/6}}\,.
\end{equation}
The term on the right-hand side only contributes to the behavior of $R_N(\xi)$ in the neighbourhood of the corresponding Stokes curve (we will later show the width of this neighbourhood to be $\mathcal{O}(\sqrt{\eps})$, and is otherwise small compared to the terms on the left-hand side. Consequently, we obtain the behavior outside of this neighbourhood by considering 
\begin{equation}\label{2.3 RNhom}
	\eps R''_N(\xi) - \chi'(\xi)^2 R_N(\xi) \sim 0,
\end{equation}
 and applying a WKB analysis to see that $R_N \sim C Z \e^{-\chi/\sqrt{\eps}}$ as $\eps \rightarrow 0$, where $C$ is an arbitrary constant. In fact, because we have specified that oscillations cannot extend ahead of the traveling wave front, we know that $C = 0$ to the left of the Stokes curve. 

To determine the behavior near the Stokes curve, we write
\begin{equation}\label{2.3 RN3}
	R_N \sim A(\xi) Z(\xi)\e^{-\chi(\xi)/\sqrt{\eps}} \qquad \mathrm{as} \qquad \eps \rightarrow 0\,,
\end{equation}
where $A(\xi)$ is a Stokes switching parameter that varies rapidly in the neighborhood of the Stokes curve. We see that \eqref{2.3 RN} and \eqref{2.3 RN3} depend directly on $\xi$, but not on $\xi \pm \lambda$.  For notational convenience, we therefore omit the arguments for $A$, $Z$, and $\chi$. We find that \eqref{2.3 RN} becomes
\begin{align}\label{this2}
	\nonumber\eps\left[\frac{A Z (\chi')^2}{\eps} - \frac{2A' Z \chi'}{\sqrt{\eps}} - \frac{2 A Z'\chi'}{\sqrt{\eps}} - \frac{A Z \chi''}{\sqrt{\eps}}\right]&\e^{-\chi/\sqrt{\eps}} \\- (\chi')^2 A Z \e^{-\chi/\sqrt{\eps}} &\sim -\frac{\eps^N (\chi')^2 Z \Gamma(2N-1/6)}{\chi^{2N-1/6}}\,.
\end{align}
Using \eqref{A.2 singulant1} and (\ref{A.2 prefactor1}), equation \eqref{this2} reduces to
\begin{align}\label{express}
	-2A'\chi'e^{-\chi/\sqrt{\eps}}  \sim -\frac{\eps^{N-1/2} (\chi')^2 \Gamma(2N -1/6)}{\chi^{2N-1/6}} \qquad \mathrm{as} \qquad \eps \rightarrow 0\,.
\end{align}
We now write \eqref{express} in terms of $\chi$ as an independent variable. Noting that $A'(\xi) = \chi'(\xi)\frac{\d A}{\d \chi}$ and rearranging, we obtain
\begin{align}\label{2.3 dAdchi}
	\diff{A}{\chi} \sim \frac{\eps^{N-1/2} \Gamma(2N -1/6)}{2\chi^{2N-1/6}} \e^{\chi/\sqrt{\eps}} \,.
\end{align}
We define polar coordinates and consider only variation that occurs in the angular direction. We thus see that
\begin{equation}
	\chi = r\e^{ i \theta}\,, \qquad \diff{}{\chi} = -\frac{ i \e^{- i \theta}}{r}\diff{}{\theta}\,.
\end{equation}
The optimal truncation in equation (\ref{2.3 truncation}) then gives
\begin{align}\label{this3}
	\diff{A}{\theta} \sim  i  r \e^{ i \theta} \frac{\eps^{r/2\sqrt{\eps}+\omega-1/2} \Gamma(r/\sqrt{\eps} + 2\omega-1/6)}{2(r \e^{ i \theta})^{r/\sqrt{\eps}+2\omega-1/6}} \e^{r\e^{ i \theta}/\sqrt{\eps}}\,.
\end{align}
We now use an asymptotic expansion of the gamma function \cite{DLMF,Abramowitz1} and expand equation \eqref{this3} and simplify to obtain
\begin{equation}\label{this4}
	\diff{A}{\theta} \sim  i \sqrt{\frac{\pi}{2}} \eps^{-1/4+1/12} r^{1/2}\exp\left(\frac{r}{\sqrt{\eps}}(\e^{ i \theta}-1) - \frac{ i \theta r}{\sqrt{\eps}} + i \theta(1+1/6+2\omega)\right)\,.
\end{equation}
The right-hand side of equation \eqref{this4} is exponentially small in $\eps$, except in the neighborhood of $\theta = 0$. We therefore define an inner region $\theta = \eps^{1/4}\bar{\theta}$ and thereby find that
\begin{equation}\label{inner_approx_noomega}
	\diff{A}{\bar{\theta}} \sim  i \sqrt{\frac{\pi}{2}} \eps^{1/12} r^{1/2} \e^{-r\bar{\theta}^2/2}\,.
\end{equation}
Consequently, by integration, we see that the behavior as the Stokes curve is crossed is 
\begin{equation}
	A \sim  i \sqrt{\frac{\pi}{2}} \eps^{1/12} r^{1/2} \int_{-\inf}^{\bar{\theta}}\e^{-r s^2/2} \d s \qquad \mathrm{as} \qquad \eps \rightarrow 0\,.\label{C:smooth}
\end{equation}
Therefore, converting back to outer coordinates, as the Stokes curve is crossed from ${\theta} < 0$ to ${\theta} > 0$, we find that
\begin{equation}
	\left[A\right]_-^+ \sim \pi i \eps^{1/12} \qquad \mathrm{as} \qquad \eps \rightarrow 0\,,,
\end{equation}
where we use the notation $[f]_-^+$ to describe the change in a function $f$ as the Stokes curve is crossed from $\theta<0$ to $\theta>0$. Therefore, the total exponentially small contribution is given in the $\eps \rightarrow 0$ limit by
\begin{equation}
	\left[R_N\right]_-^+ \sim \pi i \eps^{1/12} Z \e^{-\chi/\sqrt{\eps}} + \mathrm{c.c.}\,,
\end{equation}
where `c.c.' stands for the complex conjugate. Because $R_N(\xi)$ must be $0$ ahead of the solitary wave (i.e., when $\mathrm{Im}(\chi) < 0$), we find that behind the wave front, the exponentially small contribution is given in $\eps \rightarrow 0$ limit by
\begin{equation}
	R_N \sim \frac{\pi\mathcal{S}\Lambda i \eps^{1/12}}{\sqrt{\chi'(\xi)}}  \e^{-\chi(\xi)/\sqrt{\eps}} + \mathrm{c.c.}\,,
\end{equation}
where $\mathcal{S}$ varies rapidly from $0$ to $1$ in the neighborhood of the Stokes curve. 

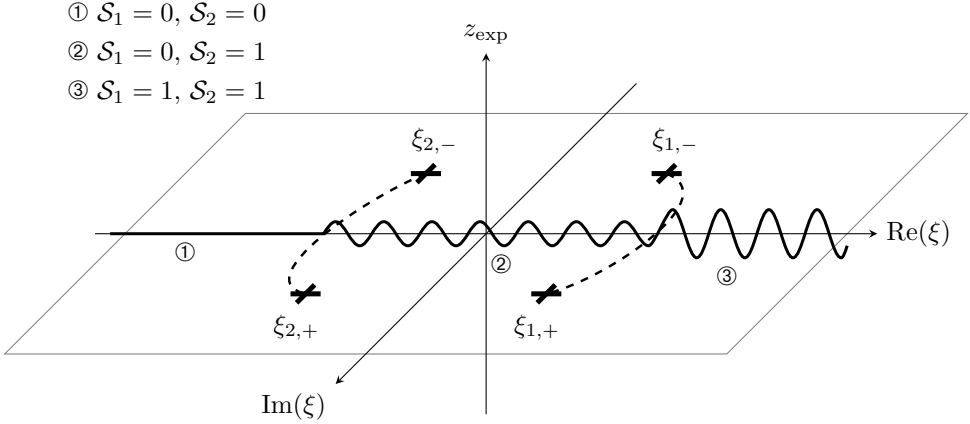
\begin{figure}
\centering
\begin{tikzpicture}
[xscale=0.8,>=stealth,yscale=0.8]

\draw[gray] (-4,2) -- (8,2) -- (4,-2) -- (-8,-2) -- cycle;
\draw[->] (2.5,2.5) -- (-2.5,-2.5) node[below left] {$\mathrm{Im}(\xi)$};
\draw[->] (-6.5,0) -- (6.5,0) node[right] {$\mathrm{Re}(\xi)$};
\draw[->] (0,-3) -- (0,3) node[above] {$z_{\mathrm{exp}}$};

\draw[line width=0.67mm] (3+0.15,1+0.15) node[above] {$\xi_{1,-}$}-- (3-0.15,1-0.15);
\draw[line width=0.67mm] (3+0.25,1) -- (3-0.25,1);

\draw[line width=0.67mm] (1+0.15,-1+0.15) -- (1-0.15,-1-0.15) node[below] {$\xi_{1,+}$} ;
\draw[line width=0.67mm] (1+0.25,-1) -- (1-0.25,-1);
\draw[line width=0.33mm,dashed] (3,1) .. controls (3+1,1-0.4) and (1+1,-1+0.24) .. (1,-1);

\draw[line width=0.67mm] (3+0.15-4,1+0.15) node[above] {$\xi_{2,-}$}-- (3-0.15-4,1-0.15);
\draw[line width=0.67mm] (3+0.25-4,1) -- (3-0.25-4,1);

\draw[line width=0.67mm] (1+0.15-4,-1+0.15) -- (1-0.15-4,-1-0.15)node[below] {$\xi_{2,+}$} ;
\draw[line width=0.67mm] (1+0.25-4,-1) -- (1-0.25-4,-1);
\draw[line width=0.33mm,dashed] (3-4,1) .. controls (3-4-1,1-0.4) and (1-4-1,-1+0.24) .. (1-4,-1);

\draw node at (-5,-0.3) {{\ding{192}}};
\draw node at (4,-0.7) {{\ding{194}}};
\draw node at (0.25,-0.5) {{\ding{193}}};

\draw[line width=0.4mm] (-6.25,0) -- (-2.7,0) sin (-2.5,0.2) cos (-2.3,0) sin (-2.1,-0.2) cos (-1.9,0) sin (-1.7,0.2) cos (-1.5,0) sin (-1.3,-0.2) cos (-1.1,0) sin (-0.9,0.2) cos (-0.7,0) sin (-0.5,-0.2) cos (-0.3,0) sin (-0.1,0.2) cos (0.1,0) sin (0.3,-0.2) cos (0.5,0) sin (0.7,0.2) cos (0.9,0) sin (1.1,-0.2) cos (1.3,0) sin (1.5,0.2) cos (1.7,0) sin (1.9,-0.2) cos (2.1,0) sin (2.3,0.2) cos (2.5,0) sin(2.7,-0.2) .. controls (2.8,-0.2) and (3,0.4) .. (3.1,0.4) cos (3.3,0) sin (3.5,-0.4) cos (3.7,0) sin (3.9,0.4) cos (4.1,0) sin (4.3,-0.4) cos (4.5,0) sin (4.7,0.4) cos (4.9,0) sin (5.1,-0.4) cos (5.3,0) sin (5.5,0.4) cos (5.7,0) sin (5.9,-0.4) cos (6,-0.2);

\node at (-3.5,3.65) [left] {\ding{192} $\mathcal{S}_1 = 0$, $\mathcal{S}_2=0$};
\node at (-3.5,3) [left] {\ding{193} $\mathcal{S}_1 = 0$, $\mathcal{S}_2=1$};
\node at (-3.5,2.35) [left] {\ding{194} $\mathcal{S}_1 = 1$, $\mathcal{S}_2=1$};

\end{tikzpicture}

\caption{Behavior of the Stokes multipliers as the Stokes curves are crossed and a schematic illustration of the associated exponentially small contribution. Recall that if $\xi \in \mathbb{R}$, then $\xi < 0$ corresponds to $\lambda p > t$, which lies in the undisturbed region; in contrast, $\xi > 0$ corresponds to $\lambda p < t$, which describes the region in the wake of the wave. In the first (undisturbed) region, both of the multipliers are $0$, and there are no oscillations. As one crosses the Stokes curve between the first and second region, $\mathcal{S}_2$ changes rapidly from $0$ to $1$. Therefore, in the second region, the Stokes contributions associated with $\xi_{2,\pm}$ are switched on. As one crosses the Stokes curve from the second region to the third, $\mathcal{S}_1$ varies rapidly from $0$ to $1$, and the contributions associated with $\xi_{1,\pm}$ are switched on, ensuring that all four contributions are present in the third region.}\label{F:StokesStruct3D}
\end{figure}

The contributions associated with $\xi_{1,-}$ and $\xi_{2,-}$ are the complex conjugates of those associated with $\xi_{1,+}$ and $\xi_{2,+}$ respectively. We therefore find that the form of the exponentially small contribution to the asymptotic behavior in the wake of the solitary wave is given by
\begin{equation}\label{2.2 zexp0}
	z_{\mathrm{exp}} \sim \left[\frac{\mathcal{S}_{1}\Lambda_{1,+} \pi i \eps^{1/12}}{\sqrt{\chi_{1,+}'(\xi)}} \e^{-\chi_{1,+}(\xi)/\sqrt{\eps}} + \frac{\mathcal{S}_{2}\Lambda_{2,+} \pi i \eps^{1/12}}{\sqrt{\chi_{2,+}'(\xi)}} \e^{-\chi_{2,+}(\xi)/\sqrt{\eps}}\right] + \mathrm{c.c.}\,, \qquad \eps \rightarrow 0\,,
\end{equation}
where $\mathcal{S}_r$ varies from $\mathcal{S}_r = 0$ on the side of the Stokes curve with $\mathrm{Im}(\chi_{r,\pm}) < 0$ to $\mathcal{S}_r = 1$ on the side of the Stokes curve with $\mathrm{Im}(\chi_{r,\pm}) > 0$. This variation occurs smoothly within a narrow neighborhood of width $\mathcal{O}(\eps^{1/4})$ about the Stokes curve. In Figure \ref{F:StokesStruct3D}, we show a schematic that summarizes the behavior of the Stokes switching coefficients $\mathcal{S}_r$.

We note that $\Lambda_{1,+}=-\Lambda_{2,+}$ and that $\chi'_{1,+}(\zeta)=\chi'_{2,+}(\zeta)$. Consequently, \eqref{2.2 zexp0} simplifies to give
\begin{equation}\label{2.2 zexp02}
	z_{\mathrm{exp}} \sim \frac{\Lambda_{1,+} \pi i \eps^{1/12}}{\sqrt{\chi_{1,+}'(\xi)}}  \left[\mathcal{S}_{1,+} \e^{-\chi_{1,+}(\xi)/\sqrt{\eps}} - \mathcal{S}_{2,+}\e^{-\chi_{2,+}(\xi)/\sqrt{\eps}}\right] + \mathrm{c.c.}\,, \qquad \eps \rightarrow 0\,,
\end{equation}

Using equations (\ref{2.3 NewToda1},\ref{2.3 NewToda2}), we also find that the exponentially small contribution to the solution for $y$ is
\begin{align}\nonumber
	y_{\mathrm{exp}} \sim &  -\frac{\eps}{2}\left[ \frac{\Lambda_{1,+} \pi i \eps^{1/12}}{\sqrt{\chi_{1,+}'(\xi+\lambda)}}  \left[\mathcal{S}_{1,+}(\xi+\lambda) \e^{-\chi_{1,+}(\xi+\lambda)/\sqrt{\eps}} - \mathcal{S}_{2,+}(\xi+\lambda)\e^{-\chi_{2,+}(\xi+\lambda)/\sqrt{\eps}}\right] \right] \\&
 -\frac{\eps}{2}\left[\frac{\Lambda_{1,+} \pi i \eps^{1/12}}{\sqrt{\chi_{1,+}'(\xi)}}  \left[\mathcal{S}_{1,+} \e^{-\chi_{1,+}(\xi)/\sqrt{\eps}} - \mathcal{S}_{2,+}\e^{-\chi_{2,+}(\xi)/\sqrt{\eps}}\right]\right] + \,  \mathrm{c.c.}\,, \quad \eps \rightarrow 0\,. \label{2.2 yexp}
\end{align}

We see from Figure \ref{F:StokesStruct3D} that the exponential asymptotic method used in the present study permits us to investigate the Stokes Phenomenon present in the solution to the diatomic Toda lattice, and we explain the origin of nanopteron behavior in terms of Stokes switching. It also permits us to isolate exponentially small effects and obtain a simple algebraic form for the nonlocal oscillations that characterize nanoptera \eqref{2.2 zexp02}.

\section{Comparison with Previous Results}\label{S:Results}

Using exponential asymptotic methods, we determined an asymptotic form \eqref{2.2 zexp02} for the exponentially small wave train. This expression is the sum of complex-conjugate terms, so we can write it as
\begin{equation}\label{2.2 zexp03}
	z_{\mathrm{exp}} \sim 2\pi\eps^{1/12}\mathrm{Re}\left( \frac{i \Lambda_{1,+}  }{\sqrt{\chi_{1,+}'(\xi)}}  \left[\mathcal{S}_{1,+} \e^{-\chi_{1,+}(\xi)/\sqrt{\eps}} - \mathcal{S}_{2,+}\e^{-\chi_{2,+}(\xi)/\sqrt{\eps}}\right] \right)\,, \qquad \eps \rightarrow 0\,.
\end{equation}

Recall that, far from the wave front, $\chi_{1,+}(\xi) \sim i \sqrt{2} \xi + k_1$ and $\chi_{2,+}(\xi) \sim i \sqrt{2}\xi + k_2$ as $\xi \rightarrow \inf$, where $k_1$ and $k_2$ are constants. Furthermore, the Stokes coefficients $\mathcal{S}$ are equal to $1$. Consequently, as $\eps \rightarrow 0$ and $\xi \rightarrow \inf$, we have
\begin{align}\nonumber
	z_{\mathrm{exp}} \sim \pi 2^{3/4}\eps^{1/12}\,\mathrm{Re}\Big( {\sqrt{i} \Lambda_{1,+}  }  \Big[&(\e^{-k_1/\sqrt{\eps}} - \e^{-k_2/\sqrt{\eps}})\cos\left(\xi\sqrt{2/\eps}\right)\\&+i(\e^{-k_1/\sqrt{\eps}} - \e^{-k_2/\sqrt{\eps}})\sin\left(\xi\sqrt{2/\eps}\right)\Big] \Big)\,. \label{2.2 zexp04}
\end{align}
Therefore, far behind the wave front, the leading-order dynamics of the exponentially small behavior in the limit that $\eps \rightarrow 0$ are described by constant-amplitude oscillations with a spatial period in $\xi$ of $\pi\sqrt{2\eps}$.

To test our results, we compare $z_{\mathrm{exp}}$ with the exact correction to the leading-order behavior, calculated in Vainchtein et al. \cite{Vainchtein1}, that we discussed in Section \ref{intro}. They derived an exact expression for corrections to the leading-order expression in terms of hypergeometric functions using an asymptotic method based on scale separation. This entailed writing an expansion
\begin{equation}\label{4:scaleseperation}
	y(n,t) \sim y_0(n,t) + \eps^2 y_c(n,t; \eps)\,,\qquad z(n,t) \sim z_0(n,t) + \eps z_c(n,t; \eps)\,,
\end{equation}
where $y_0$ and $z_0$ are the leading-order solitons given in (\ref{0:y0new},\ref{0:z0new}), and $y_c$ and $z_c$ are the full asymptotic correction terms. These correction terms were then computed exactly. Our approach above was a rather different one: we isolated the exponentially small terms and then calculated the leading-order asymptotic behavior of these terms. Vainchtein et al. \cite{Vainchtein1} found that the amplitude of the oscillations in $z(n,t)$ following the wave front to be
\begin{equation}\label{4:exact}
	\mathrm{Amplitude} \sim \frac{8 \sqrt{2} A}{\alpha \epsilon }  (\sinh\kappa)^2 \cosh \kappa \int_{0}^{\infty} \Lambda(\eta)H(\eta;\epsilon)\mathrm{d}\eta\,,
\end{equation}
where
\begin{align}
	\alpha &= \frac{\sinh\kappa}{\sqrt{2}}\,, \qquad \omega = \sqrt{\frac{2}{\eps}}\,,\qquad \nu = \frac{1}{2}\left(-1 + \sqrt{1 + \frac{16}{\epsilon}}\right)\,,\\
	A &= \sqrt{\pi}\left|\frac{\Gamma(i \omega/\alpha)\e^{-i \omega \ln 2/\alpha}}{\Gamma(\tfrac{1}{2}[\nu + 1 + \tfrac{i \omega}{\alpha}])\Gamma(-\tfrac{1}{2}[\nu - \tfrac{i \omega}{\alpha}])}\right|\,,\qquad \Lambda(\eta) = \frac{\sinh(2\eta)}{[\cosh(2\kappa) + \cosh(2\eta)]^2}\,,
\end{align}
and $H(\eta;\epsilon)$ is given in terms of the Gaussian hypergeometric function $_2 F_1$ \cite{Abramowitz1,DLMF} as
\begin{align}
	H(\eta; \epsilon) = [\cosh \eta]^{\nu + 1} (\sinh\eta) \, _2F_1\left(\frac{1}{2}\left[\nu + \frac{i\omega}{\alpha}\right] + 1,\frac{1}{2}\left[\nu - \frac{i \omega}{\alpha}\right]+1,\frac{3}{2},-(\sinh\eta)^2\right)\,.
\end{align}

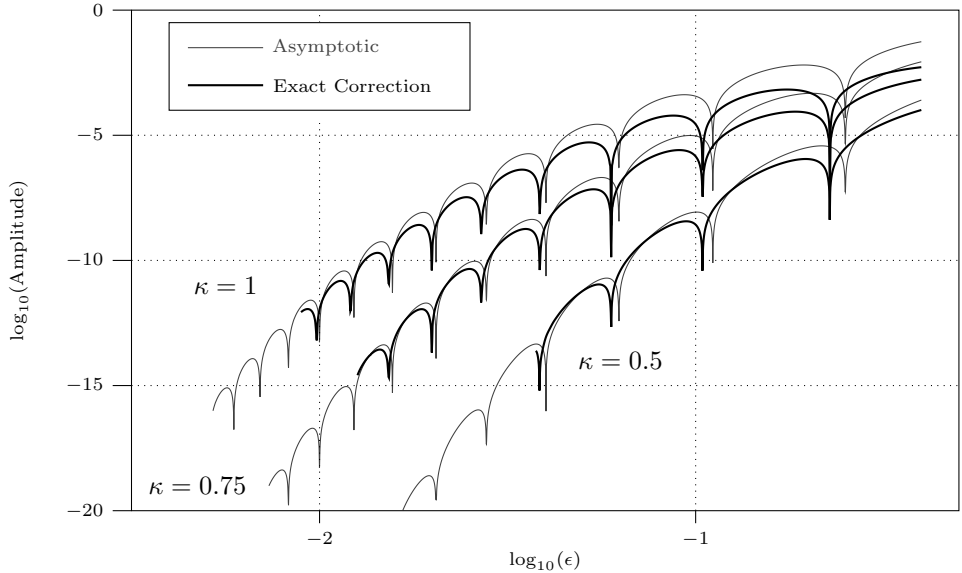
\begin{figure}
\centering
\begin{tikzpicture}
[xscale=5,>=stealth,yscale=0.333]

\draw[darkgray] plot [smooth] file {EA_k1.txt};
\draw[darkgray] plot [smooth] file {EA_k0p75.txt};
\draw[darkgray] plot [smooth] file {EA_k0p5.txt};

\draw[thick] plot [smooth] file {MS_k1.txt};
\draw[thick] plot [smooth] file {MS_k0p75.txt};
\draw[thick] plot [smooth] file {MS_k0p5.txt};

\draw[dotted] (-2.5,-5) -- (-0.3,-5);
\draw[dotted] (-2.5,-10) -- (-0.3,-10);
\draw[dotted] (-2.5,-15) -- (-0.3,-15);
\draw (-2.55,-5) node[left] {\scriptsize{$-5$}} -- (-2.5,-5);
\draw (-2.55,-10) node[left] {\scriptsize{$-10$}} -- (-2.5,-10);
\draw (-2.55,-15) node[left] {\scriptsize{$-15$}} -- (-2.5,-15);
\draw (-2.55,-0) node[left] {\scriptsize{$0$}} -- (-2.5,-0);
\draw (-2.55,-20) node[left] {\scriptsize{$-20$}} -- (-2.5,-20);
\draw (-1,-20) -- (-1,-20.5) node[below] {\scriptsize{$-1$}};
\draw (-2,-20) -- (-2,-20.5) node[below] {\scriptsize{$-2$}};

\draw[dotted] (-1,-20) -- (-1,0);
\draw[dotted] (-2,-20) -- (-2,0);

\fill[white] (-2.4,-0.5) -- (-1.6,-0.5) -- (-1.6,-4.5) -- (-2.4,-4.5) -- cycle;
\draw (-2.4,-0.5) -- (-1.6,-0.5) -- (-1.6,-4) -- (-2.4,-4) -- cycle;
\draw[darkgray]  (-2.35,-1.5) -- (-2.15,-1.5) node[right] {\scriptsize{Asymptotic}};
\draw[thick]  (-2.35,-3) -- (-2.15,-3) node[right] {\scriptsize{Exact Correction}};

\draw node at (-1.2,-14) {{$\kappa = 0.5$}};
\draw node at (-2.25,-11) {{$\kappa = 1$}};
\draw node at (-2.325,-19) {{$\kappa = 0.75$}};

\node at (-1.4,-22) {\scriptsize{$\log_{10}(\epsilon)$}};
\node[rotate=90] at (-2.8,-10) {\scriptsize{$\log_{10}(\mathrm{Amplitude})$}};

\draw(-2.5,0) -- (-2.5,-20) -- (-0.3,-20) -- (-0.3,0) -- cycle;
\end{tikzpicture}

\caption{Amplitude of the far-field oscillations given by the remainder expression $z_{\mathrm{exp}}$ in \eqref{2.2 zexp03}. The thick black curves represent the amplitude predicted by the leading-order exponentially exponentially small behavior $z_{\mathrm{exp}}$, and the thin gray curves represent the amplitude predicted by solving for the exact correction term \eqref{4:exact} \cite{Vainchtein1}. In both cases, there exist values of $\eps$ that result in complete wave cancellation (corresponding to $0$ values of the amplitude). For both the amplitude and wave-cancellation values, exponential asymptotic analysis becomes more accurate as $\eps \rightarrow 0$.
}
   \label{F:Cancellation}
\end{figure}

In Figure \ref{F:Cancellation}, we compare the far-field amplitude of the asymptotic expression in \eqref{2.2 zexp03} with a numerical computation of the expression in \eqref{4:exact}. We computed the integral in \eqref{4:exact} numerically using the inbuilt \textsc{Matlab} integral function over a sufficiently large range of $\eta$ for the result to be accurate. We exploit the fact that $\Lambda(\eta)$ decays exponentially away from some maximum point \cite{Vainchtein1}
\begin{equation}
	\eta_{\mathrm{max}} = \mathrm{arccosh}\left[\frac{\sqrt{4 + 2(\cosh\kappa)^2 + \sqrt{2}\sqrt{17+\cosh(4\kappa)} + 2(\sinh\kappa)^2}}{2\sqrt{2}}\right]\,.
\end{equation}
To obtain sufficiently high accuracy from our numerical integration, we typically integrated the expression from $\eta = 0$ to $\eta = 10\,\eta_{\mathrm{max}}$.

We observe that there is good agreement between the exact correction-term amplitude from \cite{Vainchtein1} and the amplitude \eqref{2.2 zexp03} that we obtained by calculating the asymptotic behavior. There are significant differences in the amplitude of the approximation when $\eps$ is not particularly small, thought the difference in amplitudes decreases in the limit that $\eps \rightarrow 0$. We expect this behavior, as the result obtained from the exponential asymptotic analysis in this study is a leading-order approximation of the exact correction dynamics from \cite{Vainchtein1}.  

The key differences between the present study and the work of \cite{Vainchtein1} is that we obtain the leading-order behavior of the correction $z_c$ in \eqref{4:scaleseperation}, whereas \cite{Vainchtein1} utilized the integrability of the nonlinear ordinary differential equation for $z_c$ to obtain an exact solution for this quantity in terms of a hypergeometric function. Exponential asymptotic techniques allow us to directly study the nonlocal oscillations, and they require only the exact calculation of $z_0$ and $y_0$. Such methods are therefore still applicable in problems in which the equation for $z_c$ is not integrable. Although \cite{Vainchtein1} were able to obtain an exact expression for $z_c$, we see in Figure \ref{F:Cancellation} that the simple approximation $z_{\mathrm{exp}}$ given in \eqref{2.2 zexp04} is accurate in the limit $\eps \rightarrow 0$.


\subsection{Nanopteron-Free Solutions}\label{S:nanopteronfree}

It is clear from Figure \ref{F:Cancellation} that there exist values of $\epsilon$ and $\kappa$ that produce destructive interference. Because the two oscillatory wave trains described in equation \eqref{2.2 zexp03} have identical amplitudes, it is possible to select these parameter values so that the waves precisely cancel each other out. Such configurations produce genuinely localized solitary waves, even at exponentially small orders. This behavior is consistent with a known phenomenon in diatomic granular chains \cite{Jayaprakash1, Jayaprakash2}, where a particular discrete set of parameter values (in this case, $\eps$ and $\kappa$) produce special symmetries in the system that prevent nanoptera and instead yield a family of localized solitary waves. Systems with these particular parameter values are said (for example, in \cite{Jayaprakash1, Jayaprakash2}) to satisfy an `anti-resonance condition'. 

Comparing the results in Figure \ref{F:Cancellation} from \eqref{2.2 zexp03} and \eqref{4:exact}, we see that the two methods agree well visually on the values of $\eps$ that lead to anti-resonances. Furthermore, this agreement improves significantly as $\eps \rightarrow 0$, as one would expect for an asymptotic approximation like \eqref{2.2 zexp03}.

It is possible to determine the parameter values that produce complete wave cancellation and hence nanopteron-free solutions. We know from \eqref{2.2 zexp02} that the behavior of the wave front is given by
\begin{equation}\label{2.2 zexp05}
	z_{\mathrm{exp}} \sim \frac{\Lambda_{1,+} \pi i \eps^{1/12}}{\sqrt{\chi_{1,+}'(\xi)}}  \left[ \e^{-\chi_{1,+}(\xi)/\sqrt{\eps}} - \e^{-\chi_{2,+}(\xi)/\sqrt{\eps}}\right] + \mathrm{c.c.}\,, \qquad \eps \rightarrow 0\,.
\end{equation}
To find values of $\eps$ that produce nanopteron-free solutions, we require that $z_{\mathrm{exp}}=0$ and therefore that
\begin{equation}\label{nanfree cond}
	 \e^{-\chi_{1,+}(\xi)/\sqrt{\eps}} - \e^{-\chi_{2,+}(\xi)/\sqrt{\eps}}= 0\,.
\end{equation}

Recall the integral expression in \eqref{A.2 singulant}, and note that
\begin{equation}
	\chi_{1,+}(\xi) = \int_{\xi_{1,+}}^{\xi} \chi'(s)\d s\,,\qquad \chi_{2,+}(\xi) = \int_{\xi_{2,+}}^{\xi} \chi'(s)\d s\,.
\end{equation}
We deform the contour for the latter expression to obtain
\begin{equation}
	\chi_{2,+}(\xi) = \int_{\xi_{2,+}}^{\xi_{1,+}} \chi'(s)\d s + \int_{\xi_{1,+}}^{\xi} \chi'(s)\d s = 2 \pi i +\chi_{1,+} \,,
\end{equation}
where we obtain the final expression using equation \eqref{sing2pii}. Consequently, equation \eqref{nanfree cond} becomes
\begin{equation}\label{nanfree cond2}
	 \e^{-\chi_{1,+}(\xi)/\sqrt{\eps}}(1 - \e^{-2\pi i/\sqrt{\eps}})= 0\,.
\end{equation}
Equation \eqref{nanfree cond2} holds for $\sqrt{\eps} = 1/M$, where $M$ is a positive integer. In practice, $M$ is the number of oscillations contained in the region between the two Stokes curves.

The asymptotically-predicted nanopteron-free solutions in Figure \ref{F:Cancellation} correspond to $\eps = 1/M^2$, where $M \in \mathbb{Z}_+$. This is consistent with the prediction made in \cite{Vainchtein1} that there are a countably infinite number of nanopteron-free solutions. The values of $\eps$ that produce nanopteron-free solutions do not depend on $\kappa$, as one can also see in Figure \ref{F:Cancellation}. 

In Table \ref{tableref}, we compare our prediction for the nanopteron-free values of $\epsilon$ with those calculated by \cite{Vainchtein1}. As expected, the approximation is not particularly accurate for larger values of $\epsilon$. The largest value of $\epsilon$ that satisfies the asymptotic anti-resonance condition is $\epsilon = 0.25$, which is rather different from the corresponding numerical solution of $\epsilon \approx  0.319089$. However, the error decreases significantly as $\epsilon \rightarrow 0$, as expected from an asymptotic analysis, and the predicted value is very accurate for small values of $\epsilon$.

In contrast to the asymptotic anti-resonance condition, the wave amplitude in $z_{\mathrm{exp}}$ does depend on $\kappa$. This is consistent with \cite{Vainchtein1}, who noted that the exact solution for the anti-resonance condition depends only weakly on $\kappa$, which indicates that $\kappa$ only appears as a higher-order asymptotic correction to the condition, so this dependence disappears as $\eps \rightarrow 0$.

By using exponential asymptotics, we directly studied exponentially small wave trains that occur in the solution $z_{\mathrm{exp}}$, and we derived a simple asymptotic expression \eqref{2.2 zexp04} for this behavior. This simple form yielded a new asymptotic condition ($\eps = 1/M^2$) for the existence of nanopteron-free solutions that is consistent with the analysis in \cite{Vainchtein1}. We note that \cite{Vainchtein1} used an exact correction term $z_c$ to find a condition for nanopteron-free solutions, whereas the form that we obtained using exponential asymptotics is significantly less complicated, while still capturing the correct asymptotic behavior in the limit that $\eps \rightarrow 0$ (and also giving interesting insights into the dynamics). Our approach should thus be applicable even for situations (i.e., most of the time) in which it is not possible to derive exact correction terms.

  \begin{table}
\centering
  \begin{tabular}{ | c || l | l | }
    \hline
    $M$ & Computed Value \cite{Vainchtein1} & Asymptotic Prediction: $1/M^2$ \\ \hline\hline
    2 & 0.319089 & $1/2^2 =$ 0.25 \\ \hline
    3 & 0.116993 & $1/3^2 \approx$ 0.111111 \\ \hline
    4 & 0.0633882 & $1/4^2 = $ 0.0625   \\ \hline
    5 & 0.0402105 & $1/5^2 = $ 0.04   \\ \hline
    6 & 0.0275323 & $1/6^2 \approx$ 0.0278878  \\ \hline
    7 & 0.0201695 & $1/7^2 \approx$ 0.0204082   \\ \hline
    8 & 0.0154320 & $1/8^2 =$ 0.015625  \\ \hline
  \end{tabular}

\caption{Comparison between $\eps$ values for nanopteron-free solutions calculated using the result in \cite{Vainchtein1} with $\kappa = \mathrm{arcsinh}(2\sqrt{5})$ and the asymptotic prediction obtained by solving \eqref{nanfree cond2}. The resulting anti-resonance condition gives $\eps = 1/M^2$, where $M \in \mathbb{Z}_+$ is the number of oscillations in the region between the Stokes curves.
}\label{tableref}
\end{table}


\section{Conclusions and Discussion}\label{S:Final}

We derived asymptotic solutions to singularly-perturbed, period-2 discrete particle systems governed by nearest-neighbor interactions in which the leading-order solution is a soliton. Such systems can have nanopteron solutions, which are solitary-wave solutions that are not localized, but instead possess non-decaying trains of oscillations in the wake of the wave front. These oscillations are exponentially small, so their dynamics are invisible to ordinary asymptotic power-series approaches. Using exponential asymptotic methods, we derived asymptotic descriptions of these oscillations in the singularly-perturbed limit.

We applied exponential asymptotics to determine the asymptotic behavior of solitary waves in a period-2 Toda chain, which was explored previously using different techniques in \cite{Vainchtein1}. Our asymptotic study provides a new perspective on these dynamics. In particular, we explained that the existence of the exponentially small wave train in the solution is a consequence of behavior known as the Stokes phenomenon. We found that the exponentially small oscillations are switched on in the wake of the solitary-wave front across curves known as Stokes curves, and we derived an asymptotic expression (\ref{2.2 zexp02},\ref{2.2 yexp}) for these oscillations. We compared the results of our asymptotic analysis to the results in \cite{Vainchtein1}, and we found that difference between our expressions and theirs vanishes in the asymptotic limit $\eps \rightarrow 0$.

Our investigation differs from that in \cite{Vainchtein1} in several key aspects. Using exponential asymptotics, we identified the Stokes structure of nanopteron solutions, and we explained how the exponentially small oscillatory wave trains switch on across Stokes curves. This establishes the important role that the Stokes phenomenon plays in nanopteron dynamics. The methodology that we used depends entirely on the leading-order ($\epsilon = 0$) solution, so this behavior arises without needing to compute correction terms. Although one can determine such terms for the periodic Toda lattice, as was done in \cite{Vainchtein1}, there are many systems for which such a calculation is difficult or impossible. We anticipate that the exponential asymptotic methodology in our paper will be very useful for these situations. Importantly, this approach permitted us to isolate the exponentially small terms (i.e., removing all algebraically small behavior). This produced a new, simple asymptotic approximation \eqref{2.2 zexp04} for the behavior of the oscillatory wave train.

By examining \eqref{2.2 zexp04}, we found that there exist system configurations in which the wave trains cancel entirely (see Figure \ref{F:Cancellation}). In this case, the solutions do not possess a wave train in the wake of the solitary-wave front; instead, they consist of a localized solitary wave. This is reminiscent of the results in \cite{Jayaprakash1, Kevrekidis1, Xu1}, in which special choices of wave parameters and mass ratios produce localized solitary-wave solutions due to cancellation of the oscillatory tails. In particular, \cite{Jayaprakash1} showed numerically that wave cancellation can occur in period-2 lattice systems. In our study of period-2 Toda chains, we derived a new asymptotic expression ($\eps = 1/M^2$, with $M \in \mathbb{Z}_+$) for anti-resonances, and we validated this simple formula by comparing our results to the numerical values obtained in \cite{Vainchtein1}. By isolating the exponentially small oscillatory dynamics, we were able to find a new analytical expression for an orthogonality condition that had been observed in previous work \cite{Vainchtein1}. We also demonstrated that this orthogonality condition arises as a consequence of Stokes Phenomenon, and the precise cancellation of exponentially small wave trains that appear as two different Stokes curves are crossed.

It would be interesting to generalize our analysis to study the dynamics of solutions to more-complicated period-2 particle chains. In particular, it would be fascinating to examine the woodpile system studied in \cite{Kim1} and the locally resonant granular crystal in \cite{vorot2017}. This experimental study demonstrated that solitary-wave solutions in woodpile systems take the form of nanoptera, and that there exist discrete sets of special configurations in which the oscillatory wave trains cancel to produce localized wave fronts. This suggests that an exponential asymptotic analysis of this system should reveal similar results to those that we illustrated for the Toda lattice in Figure \ref{F:Cancellation}.

Extending the results of our study to singularly-perturbed periodic particle chains of higher periodicities, such as trimers, does not require additional asymptotic techniques beyond those that we have used, although the analysis would be significantly more complicated. Such an analysis may involve a hierarchy of small parameters, depending on the mass ratios in the problem. Dynamical systems with two or more small parameters can produce a large variety of complicated behavior (see, e.g., \cite{Desroches1}), and studying nanoptera in such situations poses a fascinating challenge.

The solution that we discussed in this paper is a ``formal'' solution (i.e., we did not do a mathematically rigorous derivation of it), and it would be very interesting to provide a rigorous proof of the existence of these waves in the true solution. Recall that \cite{Faver1,Hoffman1} derived rigorous estimates for the local leading-order solitary wave, the exponentially small oscillations, and the localized remainder of a nanopteron solution of FPUT systems. We expect that one can adapt these techniques to make rigorous our statements about the existence and behavior of nanopteron solutions to the diatomic Toda lattice. 


\section*{Acknowledgements}

We thank Nalini Joshi for helpful discussions, and we thank Anna Vainchtein for useful comments. CJL was supported by Australian Laureate Fellowship Grant \#FL120100094 from the Australian Research Council.


\appendix

\section{Determining $\Lambda$}\label{S:2.2.5.2}

To determine the value of $\Lambda$, we must match the late-order term expansion in the outer region with the full local solution in an inner region near the relevant singularity. We then use Van Dyke's matching principle to match the outer expansion in the inner limit with the inner expansion in the outer limit. We perform this analysis in the neighborhood of $\xi_{1,+}$, and we then state the result for the remaining four singulants.

From equation (\ref{2.2 ansatz}), we see that the factorial-over-power ansatz breaks down when $\eps \chi^{-2} = \mathcal{O}(1)$ as $\eps \rightarrow 0$. Therefore, we define the inner scaling $\xi - \xi_s = \eps^{1/3}\overline{\xi}$. 

We must consider the local expansion of the leading-order singular behavior in a neighborhood about $\xi = \xi_s$ of radius $\mathcal{O}(\eps^{1/3})$ as $\eps \rightarrow 0$, and we then match it with the outer ansatz. As in \cite{Vainchtein1}, we see that the first nonzero correction to $y(\xi)$ is $\mathcal{O}(\eps^2)$, and the first nonzero correction to $z(\xi)$ is $\mathcal{O}(\eps)$. We therefore require that the inner correction terms be related in this fashion.

Using (\ref{z0inner})--(\ref{y01inner}), we find that the appropriate inner rescaling is
\begin{align}
	z(2p,t) &= -\log(\eps^{1/3}\overline{\xi}) + \log(1-\e^{-4\kappa}) - \log(\sqrt{2}\sinh\kappa) + \hat{z}(\overline{\xi})\,,  \label{yhat} \\
	y(2p-1,t) &= 2\log(1+\e^{-2\kappa})+ \eps \hat{y}(\overline{\xi})\,,\\
	y(2p+1,t) &= -2 \log(\eps^{1/3}\overline{\xi}) + 2\log(1-\e^{-2\kappa}) - 2 \log(\sqrt{2}\sinh\kappa)+ \eps \hat{y}(\overline{\xi}-\eps^{-1/3}\lambda)\,.
\end{align}

We write equation \eqref{2.1 Toda2b} in inner variables. At this stage, we do not required \eqref{2.1 Toda1b}, as it only affects higher-order correction terms in $\eps$ in this region that are not required for the present analysis. We retain terms up to leading order in $\eps$ and obtain
\begin{equation}\label{2.2.5 innereq}
	2\eps^{1/3}\left[-\frac{1}{2\overline{\xi}^2} + \diff{^2\hat{z}(\overline{\xi})}{\overline{\xi}^2}\right] = -\eps^{1/3}\overline{\xi}^{1/2}\beta^2\e^{\hat{z}(\overline{\xi})} + \eps^{1/3}\overline{\xi}^{1/2}\beta^2\e^{-\hat{z}(\overline{\xi})}\,.
\end{equation}
We express $\hat{z}$ in terms of the local series 
\begin{equation}\label{2.2.5 series}
	\hat{z}(\xi) \sim \sum_{j=1}^{\inf}\frac{a_j}{\xi^{j}}\,, \qquad \mathrm{as} \qquad z \rightarrow 0\,.
\end{equation}
We now expand the exponentials in equation (\ref{2.2.5 innereq}) and use the resulting relationship to determine successive values of $a_j$. With this relationship, we find that only every third term in the series beginning from $j = 3$ (i.e., $a_{3j}$ for $j \geq 1$) is nonzero. This is consistent with the local behavior of $\chi$ in equation \eqref{2.N SingulantLocal}. 

We rewrite the inner limit of the outer late-order terms \eqref{2.2.5 ansatz} as
\begin{equation}
	z_j(n,t) \sim \frac{\Lambda\Gamma(2j-{1/6})}{\beta^{1/3}({2}/{3})^{-1/6}\left[\tfrac{2\beta}{3}(\xi-\xi_s)^{3/2}\right]^{2j}}\,.
\end{equation}

\begin{figure}
\centering
\begin{tikzpicture}
[x=10,>=stealth,y=200]

\draw[thick, dashed] (22,0.544332) -- (0,0.544332);
\draw[thick, dashed] (22,0.31427) -- (0,0.31427);

\draw[black] plot [only marks, mark=*] file {LAMBDA_real.txt};
\draw[gray] plot [only marks, mark=*] file {LAMBDA_imag.txt};
\draw[thick,black] plot [only marks, mark=o] file {LAMBDA_imag.txt};
\draw[thick,black] plot [only marks, mark=o] file {LAMBDA_real.txt};
\draw (0,0) -- (22,0) -- (22,0.7) -- (0,0.7) -- cycle;

\draw[dotted] (0,0.2) -- (22,0.2);
\draw[dotted] (0,0.4) -- (22,0.4);
\draw[dotted] (0,0.6) -- (22,0.6);

\draw (0,0.2) -- (-0.5,0.2) node[left] {\scriptsize{$0.2$}};
\draw (0,0.4) -- (-0.5,0.4) node[left] {\scriptsize{$0.4$}};
\draw (0,0.6) -- (-0.5,0.6) node[left] {\scriptsize{$0.6$}};

\draw (0,0.6) -- (-0.5,0.6) node[left] {\scriptsize{$0.6$}};
\draw (0,0.6) -- (-0.5,0.6) node[left] {\scriptsize{$0.6$}};

\draw[thick] (22,0.544332) -- (22.5,0.544332) node[right] {\scriptsize{$0.544$}};
\draw[thick] (22,0.31427) -- (22.5,0.31427) node[right] {\scriptsize{$0.314$}};

\draw (0,0) -- (-0,-0.015) node[below] {\scriptsize{$0$}};
\draw (10,0.015) -- (10,-0.015) node[below] {\scriptsize{$10$}};
\draw (20,0.015) -- (20,-0.015) node[below] {\scriptsize{$20$}};

\draw[->] (22,0) -- (23,0) node[right] {$\scriptsize{j}$};

\fill (2,0.15) ellipse (0.2 and 0.01);
\fill[gray] (2,0.075) ellipse (0.2 and 0.01);
\draw[thick] (2,0.075) ellipse (0.2 and 0.01);
\draw[thick] (2,0.15) ellipse (0.2 and 0.01);

\node at (2.5,0.155)[right] {\scriptsize{Approximation for $\mathrm{Re}(\Lambda_{1,+})/(\cosh\kappa)^{1/6}$}};
\node at (2.5,0.08)[right] {\scriptsize{Approximation for $\mathrm{Im}(\Lambda_{1,+})/(\cosh\kappa)^{1/6}$}};

\end{tikzpicture}

\caption{Evaluating the ratio in \eqref{num} for increasing values of $j$. We note that this ratio converges rapidly and that the computed quantity does not depend on $\kappa$, which is factored out of the expression.}\label{F:Lambda}
\end{figure}
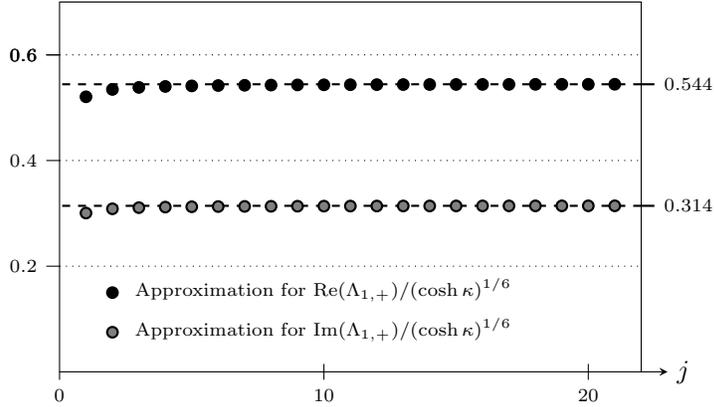

Matching the inner expansion (\ref{2.2.5 series}) with the outer ansatz (\ref{2.2.5 ansatz}) requires that
\begin{equation}\label{num}
	{\Lambda}= \lim_{j \rightarrow \inf} \frac{a_{3j}(2^{3/4} i )^{2j+1/3}(\sqrt{\cosh{\kappa}})^{2j+1/3}(2/3)^{2j-1/6}}{\Gamma(2j-1/6)}\,,
\end{equation}
where we recall that $\beta = 2^{1/4} i \sqrt{\cosh{\kappa}}$. One can numerically compute values for $a_j$ by substituting the series expression \eqref{2.2.5 series} into equation \eqref{2.2.5 innereq} and solving a recurrence relation for $a_j$ in terms of the previous series coefficients. 

One can move the term with $\kappa$ to the left-hand side of the expression \eqref{num} and evaluate $a_{3j}$ for sufficiently large values of $j$ to show that
\begin{equation}
	\frac{\Lambda_{1,+}}{({\cosh{\kappa}})^{1/6}}\approx 0.545 + 0.314i\,.
\end{equation}
We illustrate this result in Figure \ref{F:Lambda}.

A similar analysis shows that $\Lambda_{2,+} = -\Lambda_{1,+}$, and $\Lambda_{1,-}$ and $\Lambda_{2,-}$ take the corresponding complex conjugate values, giving 
\begin{equation}
	\frac{\Lambda_{1,-}}{({\cosh{\kappa}})^{1/6}}\approx 0.545 - 0.314i\,,\qquad \frac{\Lambda_{2,+}}{({\cosh{\kappa}})^{1/6}}\approx -0.545 - 0.314i\,,\qquad \frac{\Lambda_{2,-}}{({\cosh{\kappa}})^{1/6}}\approx -0.545 + 0.314i\,.
\end{equation}


\bibliography{sydrefs2.bib}
\bibliographystyle{plain}

\end{document}